\documentclass[lettersize,journal]{IEEEtran}
\usepackage{amsmath,amsfonts}
\usepackage{algorithmic}
\usepackage{algorithm}
\usepackage{array}
\usepackage[caption=false,font=normalsize,labelfont=sf,textfont=sf]{subfig}
\usepackage{textcomp}
\usepackage{stfloats}
\usepackage{url}
\usepackage{verbatim}
\usepackage{graphicx}
\usepackage{cite}
\usepackage{multirow}
\usepackage{xcolor}
\begin{document}

\title{YOLO-CCA: A Context-Based Approach\\ for Traffic Sign Detection}

\author{Linfeng Jiang,~Peidong Zhan,~Ting Bai*,~and Haoyong Yu,~\IEEEmembership{Senior Member, IEEE}
\thanks{This work was supported by the National Natural Science Foundation of China (GrantNos: 61501070, 61971078), the Natural Science Foundation of Chongqing, China (Grant No: cstc2021jcyj-msxmX0605), and Science and Technology Foundation of Chongqing Education Commission (Grant Nos: KJQN202001137, KJQN202101104).

Linfeng Jiang is with the School of Artificial Intelligence, Chongqing University of Technology, Chongqing, 404100, China. He is also with the School of Computing and College of Design and Engineering, National University of Singapore, 119077, Singapore (e-mail: linfengjiang@cqut.edu.cn).

Peidong Zhan is with the School of Artificial Intelligence, Chongqing University of Technology, Chongqing, 404100, China (e-mail:zhanpeidong@stu.cqut.edu.cn).
 
Ting Bai is with the School of Civil and Environmental Engineering, Cornell University, Ithaca, NY (email: tingbai@cornell.edu).

Haoyong Yu is with the School of Computing and College of Design and Engineering, National University of Singapore,  119077, Singapore (e-mail:
bieyhy@nus.edu.sg).
}
\thanks{}}

\markboth{IEEE TRANSACTIONS ON CIRCUITS AND SYSTEMS FOR VIDEO TECHNOLOGY}%
{Shell \MakeLowercase{\textit{et al.}}: A Sample Article Using IEEEtran.cls for IEEE Journals}

\IEEEpubid{}

\maketitle

\begin{abstract}
Traffic sign detection is crucial for improving road safety and advancing autonomous driving technologies. Due to the complexity of driving environments, traffic sign detection frequently encounters a range of challenges, including low resolution, limited feature information, and small object sizes. These challenges significantly hinder the effective extraction of features from traffic signs, resulting in false positives and false negatives in object detection. To address these challenges, it is essential to explore more efficient and accurate approaches for traffic sign detection. This paper proposes a context-based algorithm for traffic sign detection, which utilizes YOLOv7 as the baseline model. Firstly, we propose an adaptive local context feature enhancement (LCFE) module using multi-scale dilation convolution to capture potential relationships between the object and surrounding areas. This module supplements the network with additional local context information. Secondly, we propose a global context feature collection (GCFC) module to extract key location features from the entire image scene as global context information. Finally, we build a Transformer-based context collection augmentation (CCA) module to process the collected local context and global context, which achieves superior multi-level feature fusion results for YOLOv7 without bringing in additional complexity. Extensive experimental studies performed on the Tsinghua-Tencent 100K dataset show that the mAP of our method is 92.1\%. Compared with YOLOv7, our approach improves 3.9\% in mAP, while the amount of parameters is reduced by 2.7M. On the CCTSDB2021 dataset the mAP is improved by 0.9\%. These results show that our approach achieves higher detection accuracy with fewer parameters. The source code is available at \url{https://github.com/zippiest/yolo-cca}.
\end{abstract}

\begin{IEEEkeywords}
Traffic sign detection, Context, Deep learning, YOLOv7.
\end{IEEEkeywords}

\section{Introduction}
\IEEEPARstart WITH the successful application of deep learning in computer vision tasks and the continuous development of automated driving technology, the accurate detection and recognition of traffic signs has become crucial. Traffic sign detection is a key task in intelligent transportation systems~\cite{chang2012estimating}, providing vital information regarding road traffic, such as speed limits, pedestrian alerts, no honking zones. In practical scenarios, intricate environments typically introduce numerous challenges for precise traffic sign detection. Influential factors encompass illumination variations, occlusion, scale transformations, deformations, and the gradual deterioration of traffic sign visibility. In addition, certain traffic signs exhibit closely resembling appearances, making it difficult to apply traditional methods to distinguish the feature differences, thereby increasing the likelihood of classification errors.  

A number of sophisticated algorithms have been developed in the field of traffic sign detection, which are mainly categorized into two pipelines: traditional methods and deep learning methods. Early traditional methods \cite{re1,re2,re3,re4,re68,re106} typically used the color and shape characteristics of traffic signs to locate and identify. Color-based methods \cite{re4,re1} employ color information to locate image regions that potentially contain traffic signs within an image. However, these methods are highly sensitive to variations in illumination conditions and the proximity of the traffic signs. Shape-based methods \cite{re68,re106} utilize shape information to narrow down the search by filtering out areas that do not match the shape of the traffic sign. Although the shape detection methods are robust against illumination changes, it can be resource-intensive in terms of memory and computational requirements, particularly when applied to large images. Traditional traffic sign detection methods significantly rely on manual feature selection and are susceptible to limitations such as variations in lighting, deformations, and occlusions. As a result, these methods often exhibit poor detection efficiency and prediction accuracy, thus result in limited practicality in various scenarios.

In recent years, deep learning networks have been gradually applied to object detection. Unlike traditional detection methods, deep learning-based detection methods possess superior feature extraction and representation capabilities, leading to higher detection accuracy. Moreover, they are less susceptible to external environmental factors such as lighting conditions, distance variations, and occlusions. Deep learning-based object detection has now become mainstream in the field of traffic sign detection. These methods facilitate the acquisition of deeper semantic features during the training process, leading to enhanced robustness and generalization compared to handcrafted feature methods. Deep learning-based object detection algorithms can be categorized into two types according to the detection stage: (i) \textit{two-stage} and (ii) \textit{one-stage} object detection algorithms. Two-stage object detection algorithms refer to a series of algorithms based on R-CNN \cite{re6}, represented by algorithms such as Faster R-CNN \cite{re7}, Cascade RCNN \cite{re8} and Mask R-CNN \cite{re9}. Initially, these methods utilize a Region Proposal Network (RPN) to identify Regions of Interest (RoI) likely to contain objects, subsequently performing classification and localization on the RoI. One-stage detectors directly predict bounding boxes and classes on the image, including SSD-series algorithms \cite{re10,re11} and YOLO-series algorithms \cite{re12,re13,re14,re15,re16,re17}.

To achieve more accurate detection, researchers have recognized the significance of Feature Pyramid Networks (FPN) \cite{re18} in achieving high-quality results. FPN  incorporates multi-level feature fusion to address the challenge of imbalanced semantic information across different scales. This method has gained popularity due to its simplicity and efficiency, leading to its adoption in various one-stage detection models. For example, the FPN + PAN structure is used in the YOLOv7 model \cite{re17}, where the FPN structure enhances the semantic information on multiple scales by transferring the semantic features from the deeper layers to the shallower layers, while the PAN structure on the contrary transfers the localization information from the shallower layers to the deeper layers and enhances localization capabilities on multiple scales. By synergizing the advantages of different feature levels through multilevel feature fusion, better object representations can be obtained, resulting in improved detection performance.

To achieve effective multi-level feature fusion, it is crucial to minimize conflicts and inconsistencies among features from different levels. Conflicts and inconsistencies refer to obvious differences or contradictions in the representation of features at different levels\cite{re38}. When features at different levels conflict in their representations,
fusing these features may lead to confusing or inaccurate information, which may affect the final feature representation and performance. Currently, most popular mechanisms usually require complex and large parameters architectures to help achieve better multi-level feature fusion. For example, in YOLOv7\cite{re17}, the authors use extended efficient layer aggregation networks (ELAN) to enhance the extraction of effective features by fusing feature information from different feature layers using successive convolutions. However, their methods did not fully and effectively utilize the contextual information surrounding the objects to further enhance the performance of object detection. Contextual information, as a crucial clue in object detection, can provide insights into the relationships between objects and their surrounding environment. Effectively leveraging this information can assist the model in accurately identifying and localizing objects, reducing issues of omission and false detection.

Several studies \cite{re19,re20,re45,re46} have proven that including context is effective for improving object detection and other visual understanding tasks. We find that context is of great benefit for improving fusion representation over repetitive and exhaustive feature modeling. More specifically, by considering additional visual cues from a larger surrounding region as context, shallow features can be more easily resistant to visual noise, and deeper features can be more easily improved for localized detail description. As a result, gaps between features at different levels can be effectively relieved by context. Our work is inspired by the study \cite{re21}, which suggests that rich context can be decomposed into local and global contexts, with the global context being further generalized into several key features. Jointly using these two kinds of context information can avoid excessive feature processing and effectively reduce the computational cost. In contrast to the previous approaches \cite{re39,re40,re41,re42} that enhance feature representation by integrating global and local contexts, the CCA module incorporates the concept of `where' during the feature fusion process to emphasize or suppress specific features. This strategic consideration focuses on meaningful and discriminative features. The global context extracted by the CCA module takes spatial factors into account for enhanced feature fusion. The local context extracted by the CCA module is achieved through multi-scale dilated convolutions. The incorporation of local context information via multi-scale dilated convolutions enriches the feature representation. It can effectively addresses the dynamic size changes of traffic signs due to varying vehicle distances and the subsequent challenges of motion blur that arise.

In order to effectively collect and utilize context to improve multi-level feature fusion, we propose a new transformer-based context collection augmentation module (CCA) for traffic sign detection in this paper. Firstly, by incorporating features at different levels as fusion results, we first exploit rich context from the fusion results to allow the features to be better refined with the help of the Transformer. More specifically, in CCA, we try to first decompose the rich context information into local context and global context by means of adaptive Local Context Feature Enhancement (LCFE) module and Global Context Feature Collection (GCFC) module. Secondly, we extract the features of the local context and the key features of the global context as the synthesized context. Finally, we apply Transformer to process the relationship between synthesized contexts in order to identify and highlight more relevant and useful contextual information from the synthesized context for better multi-level feature fusion. The CCA is then incorporated into the one-stage detection model YOLOv7, which has better comprehensive performance. YOLO-CCA can effectively address the problem of false negatives of object detection caused by the small object sizes and false positives caused by the similarity of traffic signs. The main contributions of this paper are as follows: 

\begin{itemize}
\item{LCFE is proposed to exploit the local context in an image. We employ dilated convolutions with varying dilation rates, and subsequently utilize adaptive fusion to process the obtained feature maps with different receptive fields. This process supplements the network with valuable local context information, aiding the model in better understanding the correlation between an object and its surrounding environment. This ensures that the model maintains high precision and robustness when dealing with complex scenarios.}
\vspace{3pt}
\item{GCFC is proposed to extract the global context in an image. Global context can be understood as specific key locations within the image that assist in detecting the desired object. These key locations play a crucial role in identifying small objects that needs to be detected.}
\vspace{3pt}
\item{CCA is proposed to enhance multi-level feature fusion by exploiting contextual information. We incorporate CCA into YOLOv7 and conduct effective experiments on the TT100K and CCTSDB2021 datasets to evaluate the performance of the model. Experimental results demonstrate that the proposed method yields improved detection performance while reducing computational cost.}
\end{itemize}

The rest of the paper is organized as follows. Section~\uppercase\expandafter{\romannumeral2} reviews related works for object detection. Section \uppercase\expandafter{\romannumeral3} presents the main contributions of this work, including the implementation of the LCFE, GCFC and CCA architectures. In Section \uppercase\expandafter{\romannumeral4}, we provide experimental results and analysis. Finally, Section \uppercase\expandafter{\romannumeral5} presents concluding remarks of this work and potential revenue for future work.

\section{Related work}
\subsection{CNN-Based Traffic Sign Detection}
The emergence of CNN has significantly accelerated the development of object detection. As a widely used deep learning algorithm, CNN has a wide range of applications in the field of computer vision. The application of traffic sign detection in self-driving cars has become increasingly important. CNN-based methods are capable of automatically learning high-level semantic features with remarkable accuracy and robustness. Zhang et al \cite{re22} proposed an improved traffic sign detector based on YOLOv2 by modifying the number of convolutional layers in the network. Shirpourd et al \cite{re23} employed a combined multi-scale HOG-SVM and Faster R-CNN model to detect and recognize traffic signs inside and outside the driver's visual attention area. Yu et al \cite{re36} proposed a fusion model based on YOLOV3 and VGG19 networks that can utilize the relationship between multiple images to detect traffic signs efficiently and accurately. Chen et al \cite{re26} proposed an enhancement method based on YOLOv5, where they designed a simple cross-level loss function that assigns specific roles to each level of the model. Wang et al \cite{re241} proposed a detector for small traffic signs under multiple conditions by optimizing the detector's backbone and image enhancement network. Kamal et al \cite{re35} regarded traffic sign detection as an image segmentation problem and proposed a new network for detecting traffic signs from video sequences by merging the state-of-the-art segmentation architectures SegNet and U-Net.

In recent years, attention mechanisms have received extensive research and application\cite{re27,re28,re29,re37} in neural networks. By simulating the information processing of the human perceptual system, neural networks selectively allocate more attention to specific regions and focus on more valuable information. Zhang et al \cite{re28} improved the detection accuracy of small-scale traffic signs by combining Faster R-CNN with a channel attention mechanism to optimize the features of RoI. Wang et al \cite{re29} achieved the generation of multi-scale receptive fields and adaptive adjustment of channel features by introducing the inception structure and the channel attention mechanism, which reduced the interference of background information on detection. Gao et al \cite{re37} proposed an effective adaptive and attentive spatial feature fusion module that emphasizes or suppresses features in different regions by learning a spatial attention map.

However, the aforementioned methods predominantly focus on optimizing feature extraction through enhancements in network structures and attention mechanisms. These methods often overlook the contextual information surrounding traffic signs. This issue is particularly evident in the TT100K dataset, where certain traffic signs exhibit highly similar visual characteristics across different categories. This inter-class similarity presents a significant challenge, making it difficult for both the optimized network and the enhanced attention mechanisms to effectively guide the model in distinguishing the subtle differences between these signs. 

\subsection{Strong Baseline YOLOv7}
Recently, researchers have made significant enhancements based on the YOLOv5\cite{re15} model, resulting in the development of improved versions such as YOLOv7\cite{re17}, YOLOv8\cite{re43}, and YOLOv10\cite{re44}. In the architecture of YOLOv7, researchers introduced the SPPCSPC module as an innovative replacement for the spatial pyramid pooling fast (SPPF) used in YOLOv5 as the final layer of the backbone network. The SPPCSPC module combines the concepts of spatial pyramid pooling (SPP) and cross-stage partial networks (CSPN). By effectively integrating multi-scale features, SPPCSPC significantly enhances the adaptability of the model to objects of varying sizes.

However, in the subsequent releases of YOLOv8 and YOLOv10, researchers reverted to using SPPF. It is noteworthy that while SPPF simplifies feature representation through consecutive max-pooling operations, this process may lead to partial loss of spatial information. This issue is particularly pronounced in tasks such as traffic sign detection, which involve numerous small objects. The use of SPPF in such cases may exacerbate information loss, thereby negatively impacting detection accuracy. Given the aforementioned advantages of YOLOv7, this study adopts it as the baseline model for traffic sign detection.

\subsection{Context Modeling for Object Detection}
In object detection tasks, objects are rarely isolated entities. Instead, they often exhibit certain associations or interactions with their surrounding objects or the environment. These associations or interactions are commonly referred to as context information. How to mine the associations between them and utilize this associations to enhance the feature representation is the core problem of context information. Gong et al\cite{re32} proposed a Context-aware Convolutional Neural Network (CA-CNN) model to exploit the potential context information between objects for object detection in high-resolution remote sensing images. Chen et al\cite{re34} designed a new recursive context routing mechanism that provides a more feasible and comprehensive approach. This approach utilizes complex contexts and contextual relationships to encode contexts more efficiently. Xie et al\cite{re33} integrated valuable contextual information into 3D object detection by introducing a novel network based on VoteNet. This approach incorporates multiple layers of context information to enhance the detection and recognition of 3D objects. Leng et al\cite{re242} proposed a context-guided inference network to explore relationships between objects and use easily detected objects to help understand difficult objects. Huo et al\cite{re45} proposed a dual-branch global context module, which optimizes feature fusion by leveraging rich global context information to obtain informative feature representations.

These studies have shown that methods incorporating  contextual modeling play an significant role in improving the task of object detection. By effectively leveraging the contextual information and relationships around an object, the accuracy, robustness, and context-awareness of object detection can be improved, thereby advancing the development of object detection algorithms.
\begin{figure}[t]
\centering 
\includegraphics[width=0.45\textwidth]{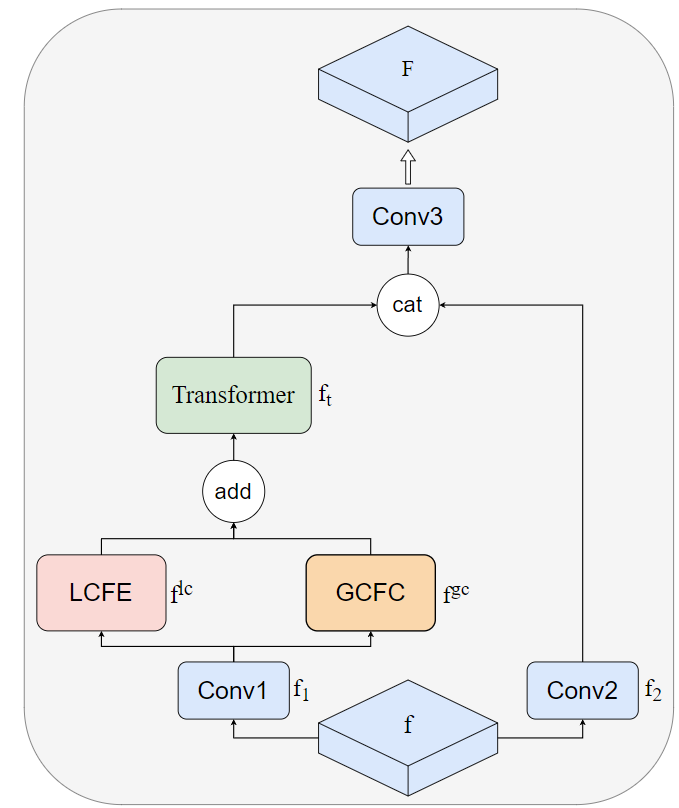}
\caption{Processing flowchart of the proposed CCA module for capturing contextual information and enhancing multilevel feature fusion. The LCFE and GCFC proposed in this paper are included in the CCA. In the figure, f represents the multilevel feature fusion result that has not been processed yet, and F represents the multilevel feature fusion result that has been refined by CCA.}
\label{fig_2}
\end{figure}
\section{Our Method}
This paper proposes a context-based traffic sign detection aalgorithm that adopts YOLOv7 as its baseline model. We introduce CCA to enhance multi-level feature fusion results, thereby improving the object detection performance of YOLOv7 without increasing computational cost. Fig. 1 shows the proposed CCA module, where LCFE is the local context feature enhancement module, GCFC is the global context feature collection module, ${f}$ is the concat-only feature maps extracted at different levels of YOLOv7 , ${F}$ is the result of CCA processing, 
${f}_{1}$ and ${f}_{2}$ represent the feature maps of ${f}$ following convolution operations Conv1 and Conv2, respectively.

Specifically, the proposed method obtains the local context and global context separately. Local context information is collected using multi-scale dilated convolution with adaptive fusion approach. Global context is generalized into several key features whose locations are predicted by an additional small neural network. The features of local context information and global generalized context are collected as synthesized context information. After combining into a synthesized context, Transformer is applied to compute the relationship between the synthesized context features in order to identify and highlight more relevant and useful context  from the synthesized context. After the Transformer computation is completed, the result of the improved feature fusion will be obtained. Then, integrate the proposed CCA into YOLOv7.

In the following sections, we  provide a detailed description of the methodology employed to extract both local and global context. Additionally, we  outline the operational formulas used in this process, and explain how the Transformer is employed to analyze the collected contexts, ultimately enhancing the results in multi-level feature fusion.

\subsection{Extraction of Local Context}

The receptive field of each layer in the convolutional neural network is fixed,  which limits its ability to capture feature information at different scales. In traffic sign images, the objects typically have simple features, insufficient texture information, and may be blurred or obscured, resulting in less extractable feature data. As a result, it becomes challenging for the model to accurately distinguish traffic signs. However, if the model is provided with local context information, such as environmental details around the traffic sign, it can make more informed inferences. For this reason, this paper proposes an adaptive local context feature enhancement module (LCFE). This module first processes the input feature map using dilated convolutions with varying expansion rates. Then it generates adaptive weights based on the processed features from different receptive fields. These features are then adaptively fused using the acquired weights. The proposed method allows the model to capture potential relationships between the traffic sign and its surrounding background or objects. By incorporating local contextual information, the inference capabilities of the network are enhanced, which enriches the feature representations and improves object detection performance.

Fig. 2 illustrates the structure of the LCFE module. The input feature map undergoes processing by dilated convolution with dilation rates 1, 2 and 3, respectively, while a 3×3 convolution is applied to capture more detailed features. This process enables the module to capture receptive fields of varying sizes, which is essential for extracting multiscale feature information.  Subsequently, outputs from these branches are adaptively fused. As illustrated in the right portion of Fig.~2, the adaptive feature fusion learns fusion weights based on input characteristics.  Different fusion weights are assigned to multi-scale features. Specifically, the process begins by concatenating the input feature map, followed by dimensionality reduction using a 1×1 convolution with a channel size of 3. A 3×3 convolution and Softmax activation are then used to map the reduced features to the corresponding weights of the three branches. Finally, the input features and spatial weights are fused through feature weighting to generate output features enriched with local context information.

\begin{figure}[t]
\centering
\includegraphics[width=0.47\textwidth]{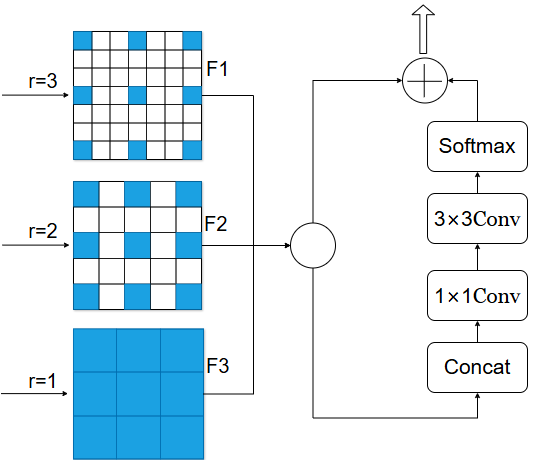}
\caption{The architecture of LCFE. The local context is extracted using multi-scale dilation convolution, and the different receptive field feature maps obtained are fused using adaptive fusion. The adaptive fusion process is shown on the right side.}
\label{fig_3}
\end{figure}
To effectively capture local context, dilated convolution kernels with dilation rates of 1, 2, and 3 are employed. The dilation rate of a dilated convolution determines the sampling step of the convolution kernel on the input, and by expanding the receptive field, a broader range of local context information can be extracted. Thus the range of local context extracted can be controlled by adjusting the dilation rate.A higher dilation rate results in a wider receptive field but may lead to the loss of detail information. Therefore, a dilation rate of 1 is retained in the LCFE module, as it is equivalent to a standard convolution operation.

We represent the extracted local context as ${f}^{lc}$, and write the formula as:
\begin{equation}
\label{deqn_ex1a}
{f}^{lc}={[DialatedConv({f}_{1})]}_{r=1,2,3},
\end{equation}
which represents the extraction fusion process of LCM. It can be further written as the following formula
\begin{equation}
\label{deqn_ex1a}
{[DialatedConv({f}_{1})]}_{r=1,2,3}={w}_{i}(F1+F2+F3),
\end{equation}
where ${F}_{1}$, ${F}_{2}$, and ${F}_{3}$ are the feature maps of dilated convolutions through three dilation rates, respectively. In addition, ${w}_{i}$ represents the fusion weights and has the following form.
\begin{equation}
\label{deqn_ex1a}
{w}_{i}=split(softmax(Conv({F}_{i})),3),
\end{equation}
where ${Conv}$ represents a 1x1 convolution operation with a channel number of 3, and ${Split( ,3)}$ represents an operation that splits the obtained feature map into three different feature maps along the channel dimension.

By utilizing the adaptive feature fusion described above, additional local context information is aggregated from multiple  receptive fields. This fusion process can enrich the information required for detection with almost no increase in computation, thereby improving detection performance for difficult-to-detect objects.

\subsection{Extraction of Global Context}
To extract the global context, key features from the entire scene are used to summarize and represent useful information. Global context can be defined as the identification of key locations within an image. When detecting traffic signs, key locations such as the ends of roads, intersections, turns, or special areas like schools, hospitals, construction zones can serve as critical positions in the global context. Enhancing the features at these key locations can aid in the detection of traffic signs. To identify key features within a scene, a neural network is designed to autonomously learn and determine which features are most relevant. The extraction process is divided into two steps: (i) a small network is used to predict the locations of key features; (ii) key features are collected as a generalized global context based on the predicted locations. The specific process is illustrated in Fig. 3, where the feature map is first passed through a convolutional layer with a convolutional kernel size of 1×1 and a number of output channels of 4 (we assume  that four key features are predicted). Then, the maximum value of each channel is determined using a global maximum pooling operation. This operation returns the location $(x, y)$ and the maximum value of the channel, which serves as the importance score. Subsequently, the collected features are highlighted and enhanced based on the feature map before GCFC processing as well as the predicted coordinate locations and scores.
\begin{figure}[t]
\centering
\includegraphics[width=0.36\textwidth]{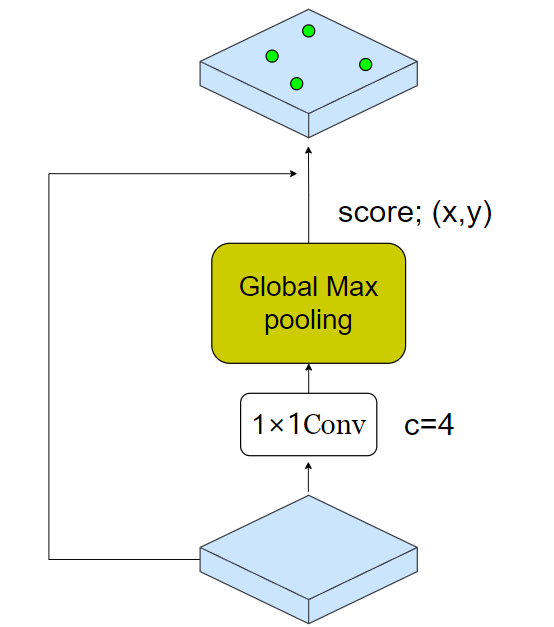}
\caption{The architecture of GCFC. The four green dots in the figure represent the key locations of the four global contexts extracted.}
\label{fig_4}
\end{figure}
If ${f}^{gc}$ is used to denote the desired global context. Given a set of locations $p=\left \{(x,y)_{1}, ...,(x,y)_{n} \right \} $ is used to represent the predicted locations of key global features, then the generalized global context is obtained according to the following formula: ${f}^{gc}$ .
\begin{equation}
\label{deqn_ex1a}
{f}^{gc}=\mu \left ( {f}_{1} , p\right ),
\end{equation}
where $\mu \left ( {f}_{1} , p\right )$ is a function that collects features from the input feature map ${f}_{1}$ in the set of locations, and is defined as:
\begin{equation}
\label{deqn_ex1a}
\mu \left ( {f}_{1} , p\right )=\left [ \left ( {f}_{1}, {\left (x,y \right )}_{1}\right ),..., \left ( {f}_{1}, {\left (x,y \right )}_{n}\right )\right ],
\end{equation}
where $\left ( {f}_{1}, {\left (x,y \right )}_{n}\right )$ is to collect detailed visual features at position $ {\left (x,y \right )}_{n}$.
We establish a network model to automatically predict the position set. To accomplish this task, an importance score is introduced for the locations of the feature map. Assuming that $n$ key global contextual features are collected, then the importance score has $n$ dimensions. Here the importance score is represented as ${S}_{i}$ $i=1,2,...,n$ where ${S}_{n}$ represents the $n^{th}$ importance score map. In GCFC , we use convolution to calculate ${S}_{i}$.
\begin{equation}
\label{deqn_ex1a}
{S}_{i}=split\left ( Conv\left ( {f}_{1}\right ),n\right ),
\end{equation}
where $Conv$ represents a $1x1$ convolution operation with n channels, and $Split( ,n)$ represents an operation that splits the obtained feature map into n different feature maps along the channel dimension. Each segmented feature map ${S}_{n}$ is 1-dimensional and can be considered as an importance score map for the desired $n^{th}$ key global context feature location. By computing and splitting the importance score map, we collect P whose corresponding importance score has the highest value in its associated score map. For example, when collecting the location of the ${n}^{th}$ key global context feature, we perform a global maximum pooling operation on the feature map ${S}_{n}$ to help recognize the location $ {\left (x,y \right )}_{n}$  that should receive attention and the corresponding maximum importance   ${score}_{n}$.
\begin{equation}
\label{deqn_ex1a}
{score}_{n}; {\left ( x,y\right )}_{n}=Maxpool({S}_{n}),
\end{equation}
where ${Maxpool}$ is the global maximum pooling operation. The ${score}_{n}$ with sigmoid compression is multiplied with the corresponding features and further the operation $\left ( {f}_{1}, {\left (x,y \right )}_{n}\right )$ is denoted as:
\begin{equation}
\label{deqn_ex1a}
\left ( {f}_{1}, {\left (x,y \right )}_{n}\right )={{f}_{1}}_{{\left ( x,y\right )}_{n}}\times sigmoid\left ({score}_{n} \right ),
\end{equation}
where ${{f}_{1}}_{{\left ( x,y\right )}_{n}}$ represents a specific feature located at $ {\left (x,y \right )}_{n}$ on the feature map ${f}_{1}$.

By constructing the GCFC as described above, the key features that serve as the global context are extracted, which improves the detection performance of small objects.
\begin{figure}[]
\centering
\includegraphics[width=0.45\textwidth]{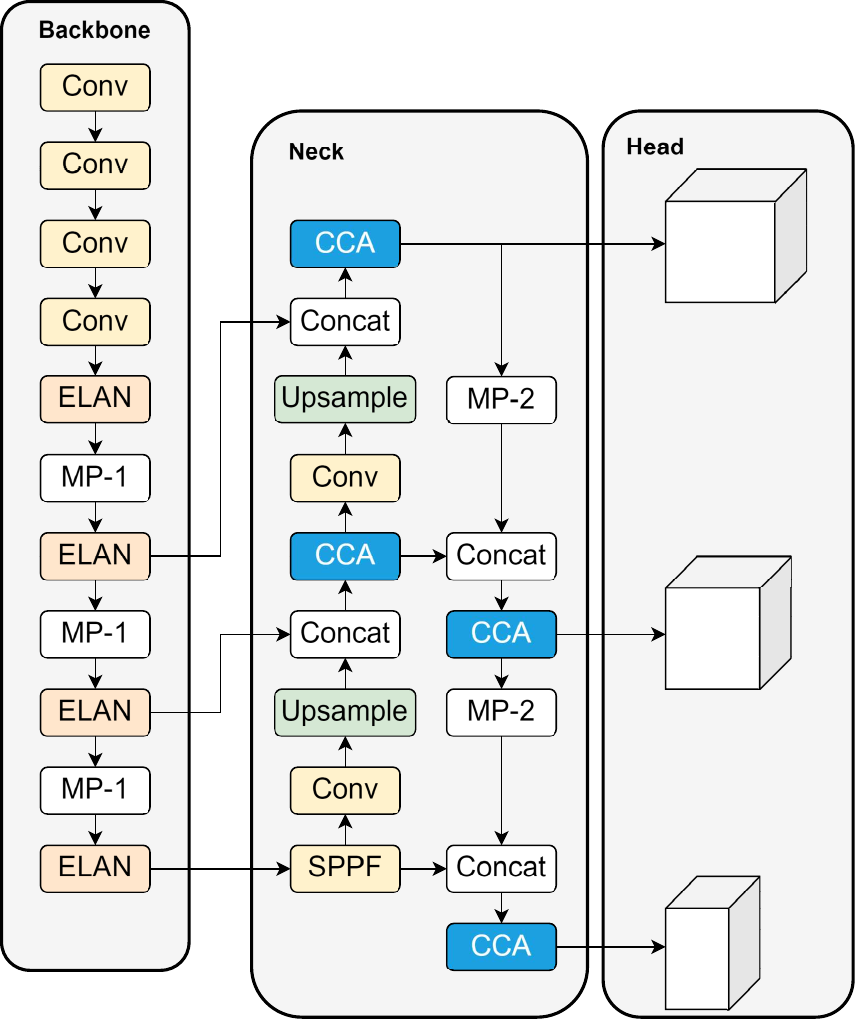}
\caption{The diagram of the proposed YOLO-CCA
model.}
\label{fig_4}
\end{figure}
\subsection{CCA}
In this paper, a new transformer based context  enhancement module is proposed to improve the results of multi-level feature fusion for effective object detection. The proposed module can enhance detection accuracy while maintaining computational efficiency.  As illustrated in Fig. 1, the feature map $f$ is first passed through Conv1 and Conv2, which are convolutional layers with 3×3 kernel size and stride size of 2, The primary function of these layers is to reduce the number of channels in the feature map $f$, thereby decreasing computational complexity. Then the feature maps passing through Conv1 are extracted to collect local context information and global context key features through LCFE module and GCFC module ,respectively. Then the two kinds of contexts are spliced to get the synthesized context. The set of synthesized context information can be represented as:
\begin{equation}
\label{deqn_ex1a}
Q\left ( {f}_{1}\right )={f}^{lc}+{f}^{gc}.
\end{equation}

Following the synthesis of the contextual information, the next step is to put the context information into the transformer for refining. Transformer is highly effective at combing complex relationships between features, making them well-suited for transforming the synthesized context into a more robust representation, thereby enhancing the overall feature fusion process.

However, the decoder part of the transformer model structure is not used because adding the decoder part will increase the extra overhead of the CCA model. Once the transformer refines the synthesized context, it is then spliced with the feature maps obtained through Conv2. The obtained feature maps are subsequently passed through Conv3 to get the final fusion result, which is a convolutional layer with a convolutional kernel size of 3×3. The mathematical formulation of CCA is shown below
\begin{equation}
\label{deqn_ex1a}
{f}_{t}=Trans\left ( q,k,v=Q\left ( {f}_{1}\right )\right ), 
\end{equation}
\begin{equation}
\label{deqn_ex1a}
 F=Conv3\left ({f}_{t}+{f}_{2} \right ),
\end{equation}
where ${Trans}$ refers to the Transformer module, and $q$, $k$, and $v$ represent query, key, and value in the Transformer, respectively.
\subsection{YOLO-CCA}
\begin{figure}[t]
\centering
\includegraphics[width=0.43\textwidth]{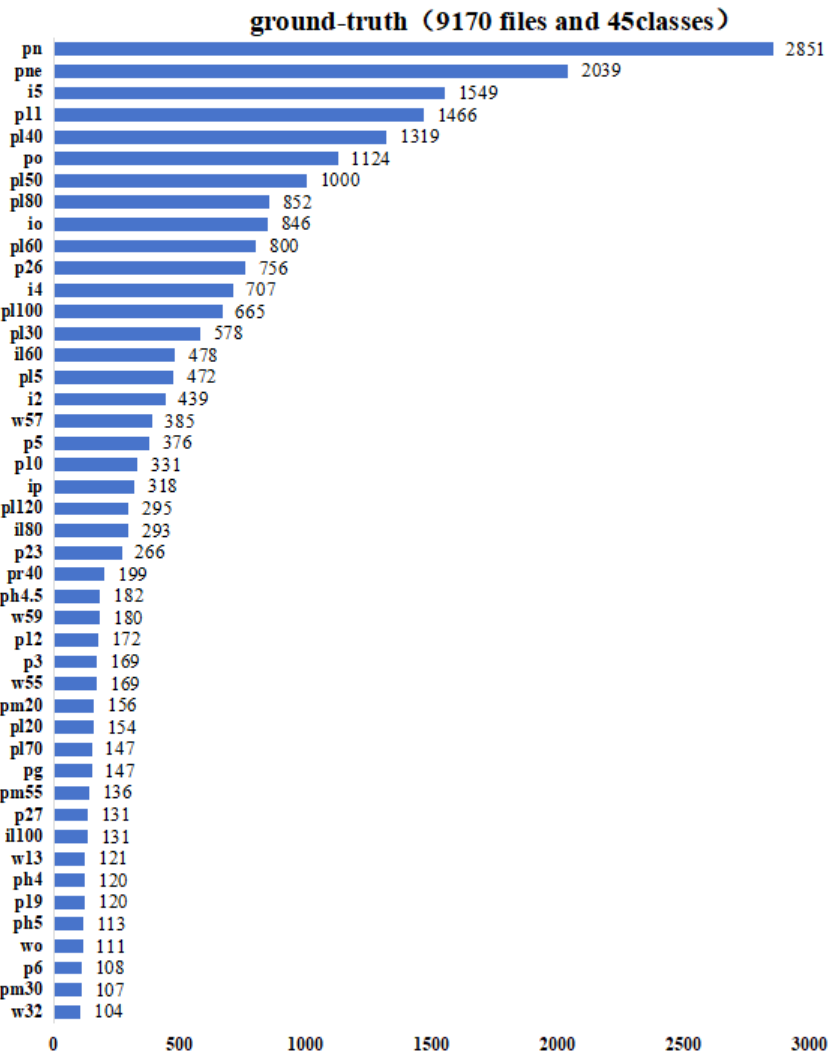}
\caption{Statistical chart of the number of classes in TT100K.}
\label{fig_5}
\end{figure}
\begin{figure}[h]
\centering
\includegraphics[width=0.45\textwidth]{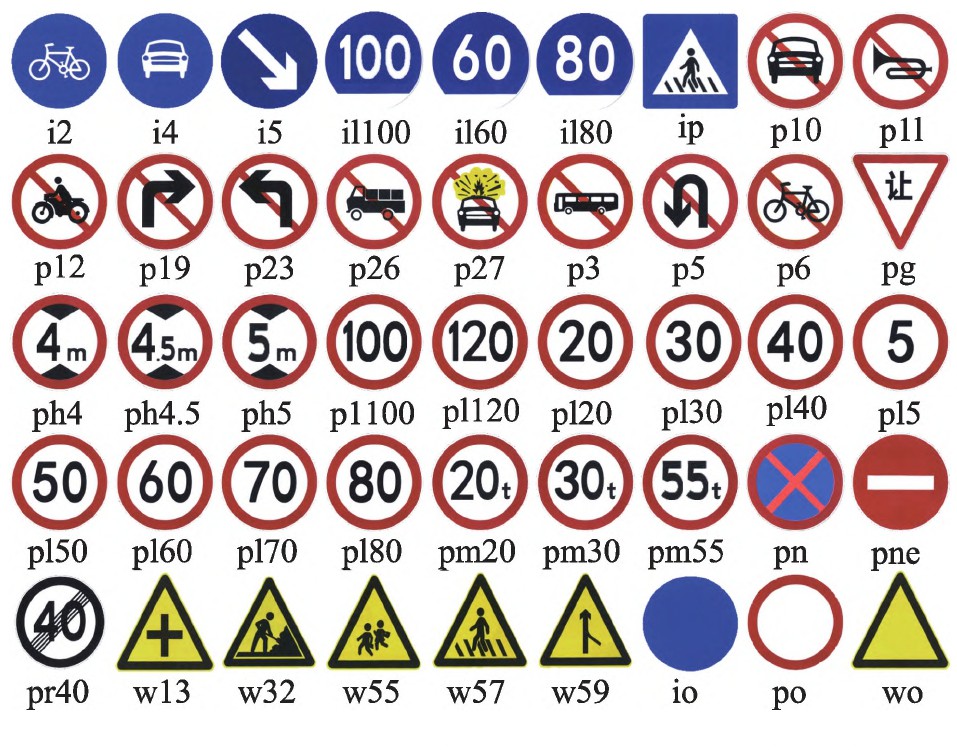}
\caption{45 classes participating in the evaluation on the TT100K dataset.}
\label{fig_6}
\end{figure}
YOLOv7 uses Extended Efficient Layer Aggregation Network (ELAN) to enhance the extraction of effective features by feature fusion of feature information from different feature levels through successive convolution. However, this approach introduces a significant amount of computational complexity. In this work, we integrate CCA into YOLOv7 to enhance multi-level feature fusion and significantly alleviate the inconsistency between features. The architecture  of YOLO-CCA is shown in Fig 4. Specifically, in YOLOv7, the four ELAN sections of the neck are replaced with CCA. This modification not only improves the model's performance but also reduces the number of training parameters required.

\section{Experiments AND Results}
\subsection{Datasets}
In this paper, we validated the effectiveness of the algorithm on the TT100K and CCTSDB2021 datasets.
\subsubsection{TT100K}
The traffic sign dataset utilized in the experiments is TT100K, jointly developed by Tsinghua University and Tencent. The dataset comprises images with a resolution of 2048×2048 pixels. Nearly half of the classes in the original dataset have single-digit instances, which makes the data distribution severely unbalanced. Therefore, only 45 classes with more than 100 instances are retained in the TT100k dataset for the experiment. Following  proportional splitting, the  training and validation sets contain 7336 and 1834 images, respectively. Fig. 5 shows the statistics of the number of samples per class for the TT100K dataset. Fig.~6 shows the 45 classes participating in the evaluation on the TT100K dataset. ‘i' represents traffic signs related to instructions, `p' represents traffic signs related to prohibitions, and `w' represents traffic signs related to warnings.

Traffic signs are classified into three classes based on the size of the pixels occupied by the instances of the traffic signs. The pixel interval occupied in [(0×0),(32×32)] is small object, [(32×32),(96×96)] is medium object, and [(96×96),(400×400)] is large object. The specific size distribution is shown in Fig. 7. It can be seen that the TT100K dataset contains a large number of small-object traffic signs, which means that the TT100K dataset is suitable for testing the model's effectiveness for small-object detection of traffic signs.

\begin{figure}[t]
\centering
\includegraphics[width=0.46\textwidth]{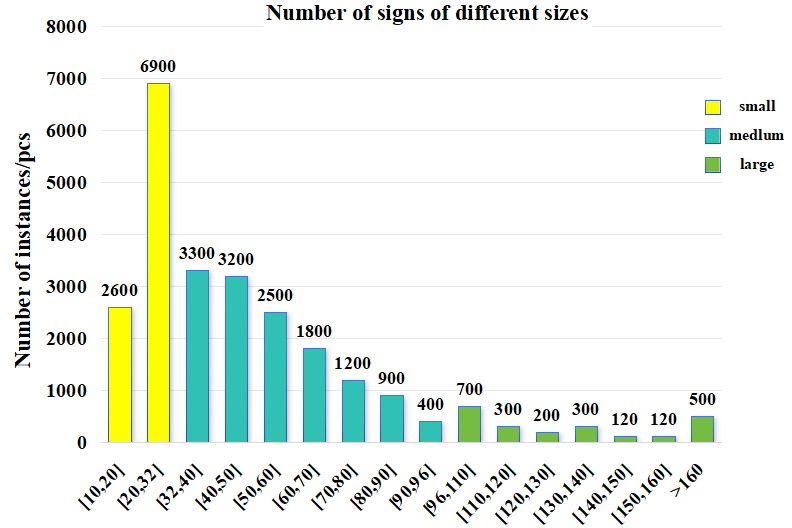}
\caption{Instance size distribution in the TT100K Dataset.}
\label{fig_7}
\end{figure}
\begin{table}[]
\caption {Experimental hardware and software configuration.}
\label{tab:table1}
\centering
\begin{tabular}{ll}
\hline
\textit Environment  & Configuratio Inforemation \\ \hline
GPU                  & TITAN RTX                 \\ 
Memory               & 32G                       \\ 
Operating ystem      & Ubuntu 20.04              \\ 
Hard disk            & 2TB                       \\ 
Programming          & PyTorch 1.11; Python 3.9  \\ \hline
\end{tabular}
\end{table}
\subsubsection{CCTSDB2021}
CCTSDB2021 is a rich dataset with professional shooting angles, which is closer to the actual traffic scene. CCTSDB2021 adds more than 4000 real traffic scene images and corresponding lables to CCTSDB2017 and replaces many of the original easy-to-detect images with difficult samples to adapt to the complex and changing detection environment. The CCTSDB2021 dataset includes pictures of traffic signs under four different weather conditions, such as rain, fog, cloudy and snow, as well as pictures of traffic signs under dim environment at night. The data categories are divided into 3 classes: mandatory signs, prohibition signs and warning signs, with 16356 images in the training set and 1500 images in the test set.

\subsection{Experimental Environment and Parameters}
The experimental hardware and software configuration for this study are shown in Table \uppercase\expandafter{\romannumeral1}. In the process of training, the initial value of the learning rate was 0.01, and we used the cosine annealing strategy to reduce the learning rate. The model had an initial input size of 640×640. The training was performed for 300 epochs with a batch size of 2.

\begin{table*}[htb]
\caption {Ablation analysis for the proposed method on the TT100K dataset.}
\label{tab:table1}
\centering
\begin{tabular}{l p{60pt}<{\centering}p{55 pt}<{\centering}p{55pt}<{\centering}p{55 pt}<{\centering}p{55 pt}<{\centering}p{55 pt}<{\centering}}
\hline
\textbf{Experiment} & \textbf{Model} & \textbf{Params} & \textbf{P} & \textbf{R} & \textbf{mAP@.5} & \textbf{mAP@.5:.95} \\ \hline
\multicolumn{1}{l}{A}                   & YOLOv7         & 36.5M           & 0.855      & 0.813      & 0.882           & 0.696               \\ 
\multicolumn{1}{l}{B}                   & YOLOv7+Trans   & \textcolor{blue}{32.0M}           & 0.836      & 0.852      & 0.898           & 0.702               \\ 
\multicolumn{1}{l}{C}                   & B+LCM          & 33.8M           & 0.847      &\textcolor{blue}{0.881}       & 0.916           & 0.713               \\ 
\multicolumn{1}{l}{D}                   & B+GCM          & 32.9M           & 0.840      & 0.862      & 0.903           & 0.704               \\ 
\multicolumn{1}{l}{E}                   & Ours           & 33.8M           & \textcolor{blue}{0.863}      & 0.880      &\textcolor{blue}{0.921}            &\textcolor{blue}{0.730}               \\ \hline
\end{tabular}
\hfil
\end{table*}
\begin{table*}[htb]
\caption {the quantitative comparison of different methods on the TT100K dataset.}
\label{tab:table1}
\centering
\begin{tabular}{l p{55pt}<{\centering}p{55 pt}<{\centering}p{55pt}<{\centering}p{55 pt}<{\centering}p{55 pt}<{\centering}p{55 pt}<{\centering}}
\hline
\textbf{Model} & \textbf{Params} & \textbf{FLOPS} & \textbf{P}     & \textbf{R}     & \textbf{mAP@.5} & \textbf{mAP@.5:.95} \\ \hline
\multicolumn{1}{l}{Faster-RCNN}    & 41.6M           & 211.5G         & 0.687          & 0.641          & 0.706           & 0.569               \\
\multicolumn{1}{l}{YOLOv3}         & 58.8M           & 156.0G         & 0.633          & 0.713          & 0.717           & 0.558               \\
\multicolumn{1}{l}{YOLOv3-SPP}     & 59.9M           & 156.9G         & 0.758          & 0.749          & 0.802           & 0.628               \\
\multicolumn{1}{l}{YOLOv5s}        & 6.8M           & 16.3G           & 0.840          & 0.778          & 0.851           & 0.663               \\
\multicolumn{1}{l}{YOLOv5l}        & 44.2M           & 109.0G         & 0.825          & 0.780          & 0.856           & 0.674               \\
\multicolumn{1}{l}{YOLOv8n}        & 2.8M  & \textcolor{blue}{8.2G}  & 0.730          & 0.690          & 0.751           & 0.565               \\
\multicolumn{1}{l}{YOLOv8s}        & 10.6M           & 28.7G          & 0.848          & 0.766          & 0.854           & 0.661               \\
\multicolumn{1}{l}{YOLOv8m}        & 24.6M           & 79.2G          & 0.876         & 0.809          & 0.888           & 0.693               \\
\multicolumn{1}{l}{YOLOv9}   & 29.9M & 118.3G & \textcolor{blue}{0.898} & 0.769 & 0.885 & 0.691 \\
\multicolumn{1}{l}{YOLOv10n} & \textcolor{blue}{2.6M}  & 8.5G   & 0.752 & 0.670 & 0.749 & 0.579 \\
\multicolumn{1}{l}{YOLOv10s} & 7.7M  & 25.0G  & 0.816 & 0.775 & 0.843 & 0.659 \\
\multicolumn{1}{l}{YOLOv10m} & 15.7M & 64.3G  & 0.852 & 0.786 & 0.862 & 0.681 \\
\multicolumn{1}{l}{YOLOv10b} & 19.5M & 99.1G  & 0.887 & 0.779 & 0.875 & 0.692 \\
\multicolumn{1}{l}{YOLOv10l} & 24.6M & 127.6G & 0.868  & 0.784 & 0.876 & 0.694  \\
\multicolumn{1}{l}{YOLOv7(baseline)}         & 36.5M           & 105.4G         & 0.855          & 0.813          & 0.882           & 0.696               \\
\multicolumn{1}{l}{Ours}           & 33.8M           & 99.2G          & 0.863 & \textcolor{blue}{0.880} & \textcolor{blue}{0.921}  & \textcolor{blue}{0.730}      \\ \hline
\end{tabular}
\end{table*}
\subsection{Evaluation Metrics}
We adopt the commonly used mean Average Precision (mAP) metric in object detection tasks. True Positive (TP) denotes that the detected traffic sign is correct. False Positive (FP) denotes that the detected traffic sign is wrong. False Negative (FN) denotes that the missed traffic sign is detected. When IoU is greater than or equal to the threshold, the prediction box is considered True Positive, otherwise False Positive. Precision (P) mainly measures the degree of misdetection of the model, and Recall (R) mainly measures the degree of omission of the model, which is calculated as

\begin{equation}
\label{deqn_ex1a}
 P=\frac{TP}{TP+FP},  
\end{equation}

\begin{figure*}[]
\centering
\begin{minipage}{\linewidth}
  \centering
\end{minipage}
\vspace{0.4em}
\begin{minipage}{\linewidth}
  \centering
  \includegraphics[width=2.3in]{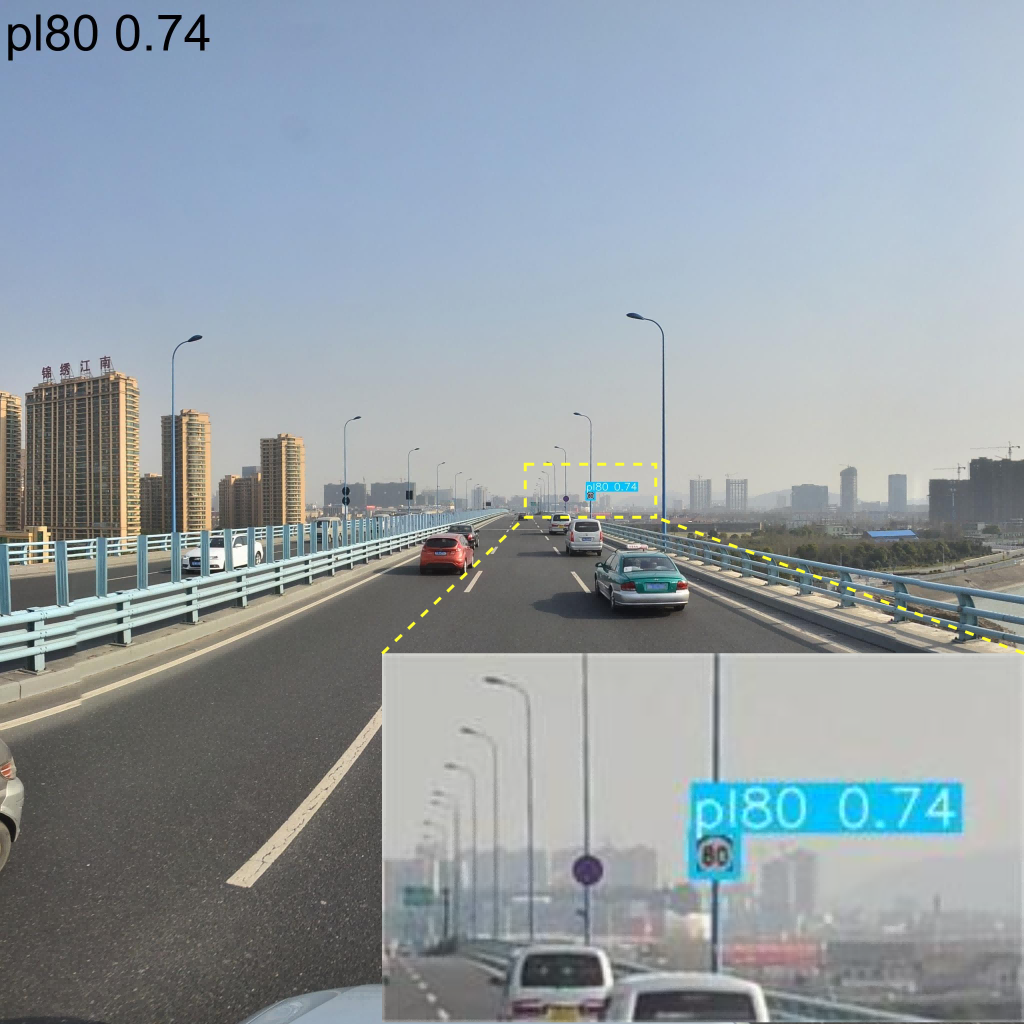}
  \includegraphics[width=2.3in]{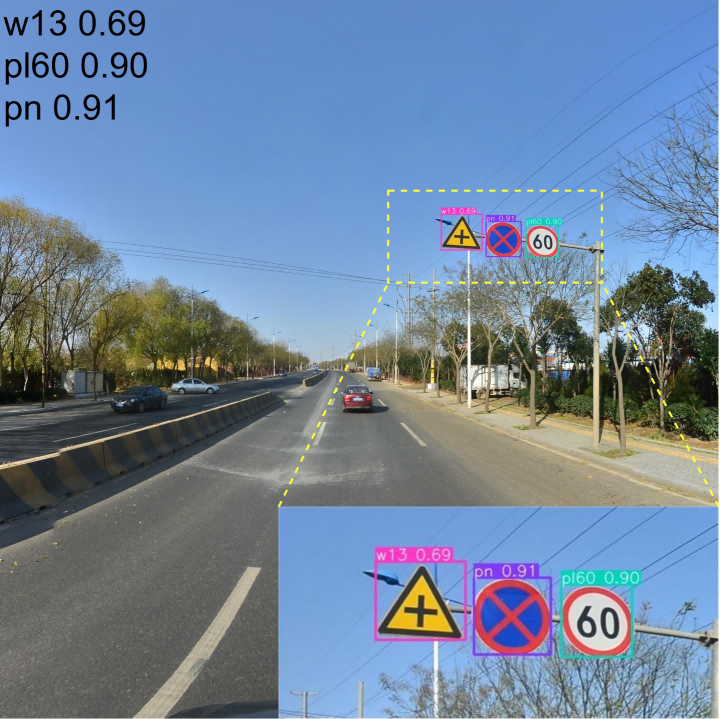}
  \includegraphics[width=2.3in]{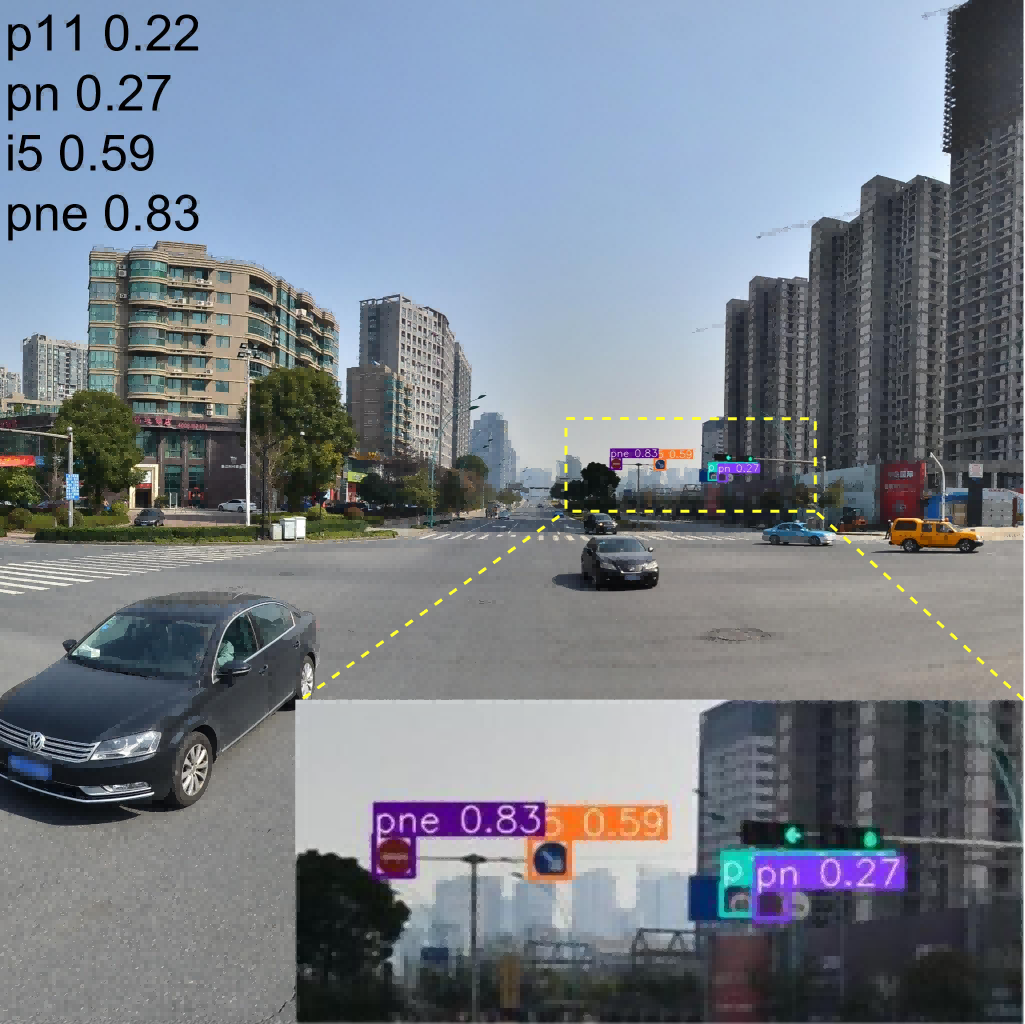}
\end{minipage}
\vspace{0.4em}
\begin{minipage}{\linewidth}
  \centering
  \includegraphics[width=2.3in]{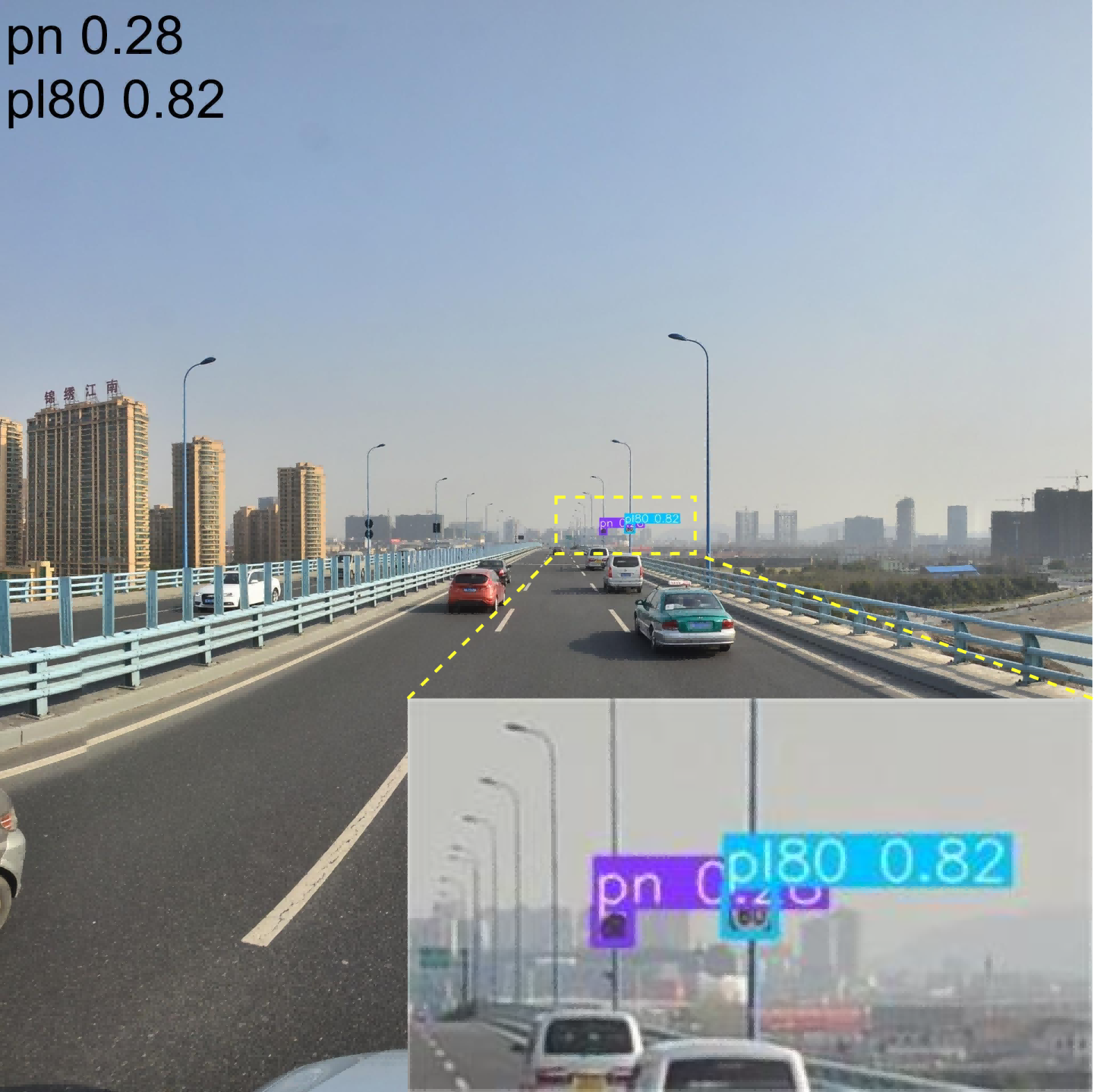}
  \includegraphics[width=2.3in]{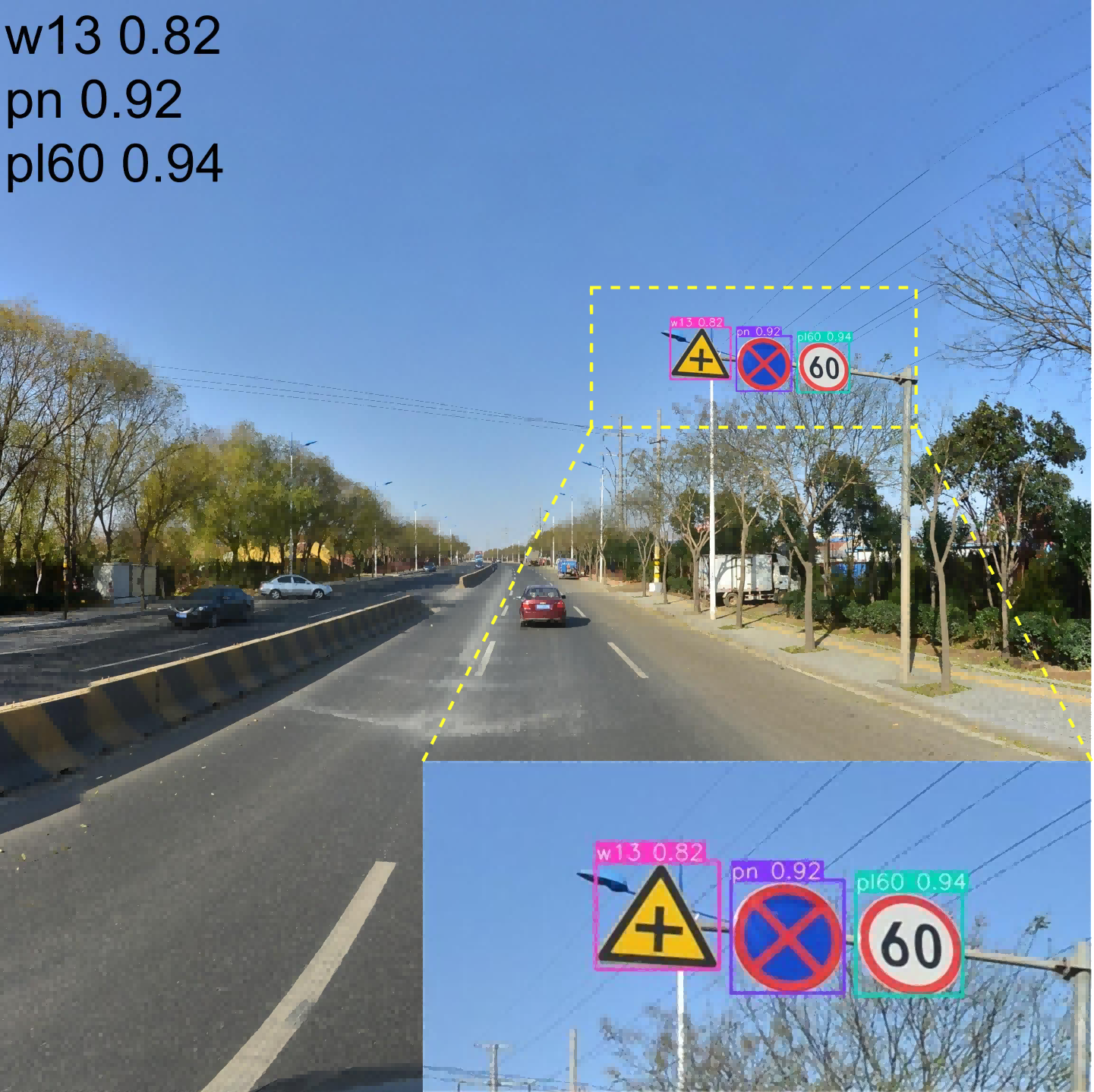}
  \includegraphics[width=2.3in]{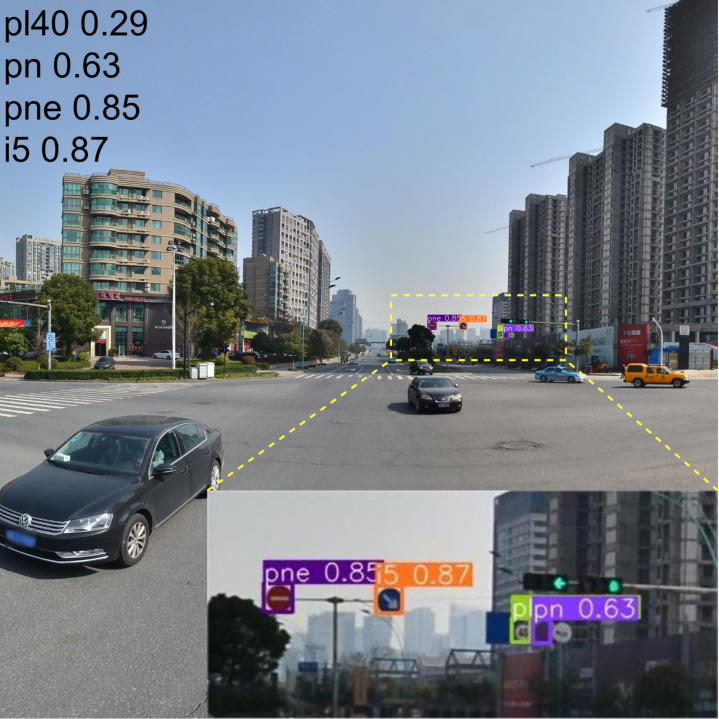}
\end{minipage}
\caption{Detection examples on the TT00K dataset. The first row displays the detection baseline. The second row displays the detection of the proposed YOLO-CCA, where all the traffic signs in the second row are accurately detected, and YOLO-CCA also outperforms the baseline model in terms of the confidence of recognizing the traffic signs. The area containing the traffic sign is enlarged and placed below the images. Reprint the detection class and the confidence score in the upper-left corner of the images. The TT100K dataset contains 45 different categories of traffic signs, with the names of each category and their corresponding images shown in Fig. 6. For example, in the detection image in the bottom left corner, `pn' represents a no parking sign, while `pl80' indicates a speed limit of 80 km/h.}
\label{fig_8}
\end{figure*}

\begin{equation}
\label{deqn_ex1a}
 R=\frac{TP}{TP+FN}.  
\end{equation}

Average Precision (AP) denotes the area enclosed by the Precision-Recall (P-R) curve with the axes, and is a comprehensive metric that combines precision and recall to comprehensively evaluate the object detection model. The mAP is the mean value of the APs of all the classes, and the larger the mAP value is, the better the detection effect is. The APs and mAPs are calculated by. 

\begin{equation}
\label{deqn_ex1a}
AP=\int_{0}^{1}PdR, 
\end{equation}

\begin{equation}
\label{deqn_ex1a}
mAP=\frac{1}{N}\displaystyle\sum_{i=1}^{N}{AP}_{i},
\end{equation}
where $N$ represents the number of classes and ${AP}_{i}$ represents the average precision of the $i$th class.

In addition, we compare the mAP at different IoU thresholds. The mAP@.5 is used to evaluate the mAP at an IoU threshold of 0.5. The mAP@.5:.95 represents the mean mAP across IoU thresholds from 0.5 to 0.95, which represents a stricter evaluation metric. To further assess the model's size and speed, we also evaluated the model's parameter count (Params) and the number of floating-point operations per second (FLOPs).

\begin{figure*}[h]
\centering

\subfloat[{\fontsize{8pt}{8pt}\selectfont Foggy day test comparison results}]{\includegraphics[width=2.6in]{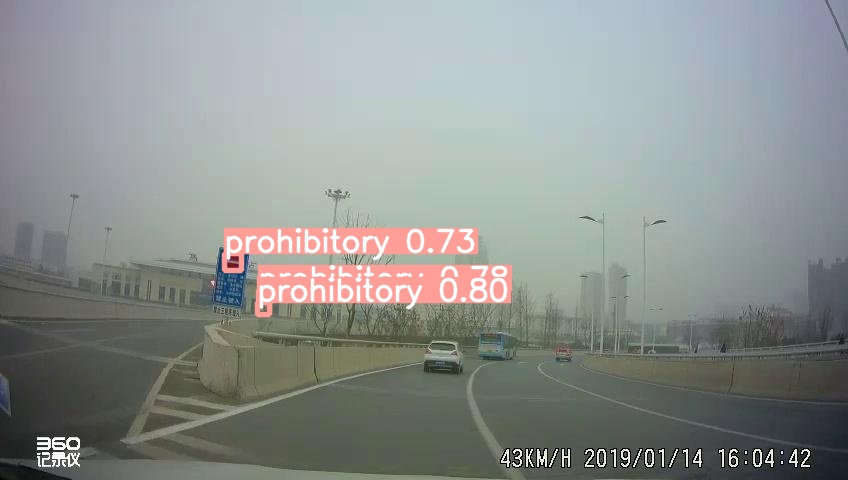}
              \vspace{0.4em}
              \includegraphics[width=2.6in]{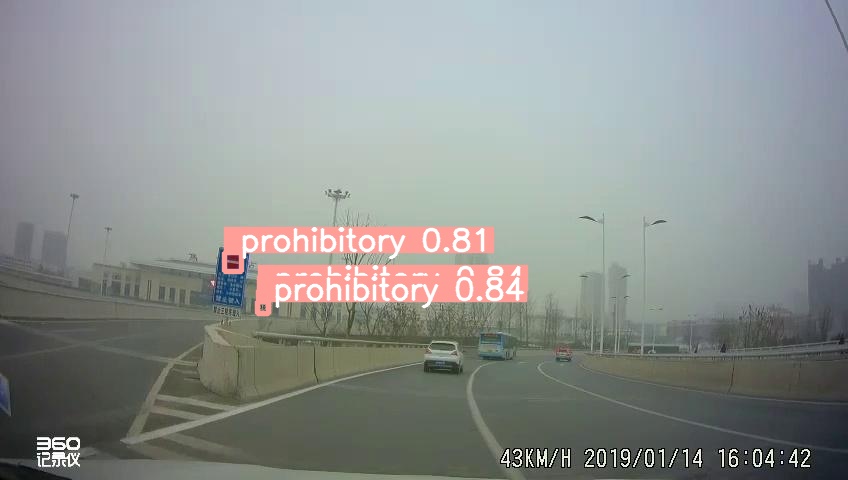}
              \label{fig_first_case}}
\hfil
\subfloat[{\fontsize{8pt}{8pt}\selectfont Rainy day test comparison results}]{\includegraphics[width=2.6in]{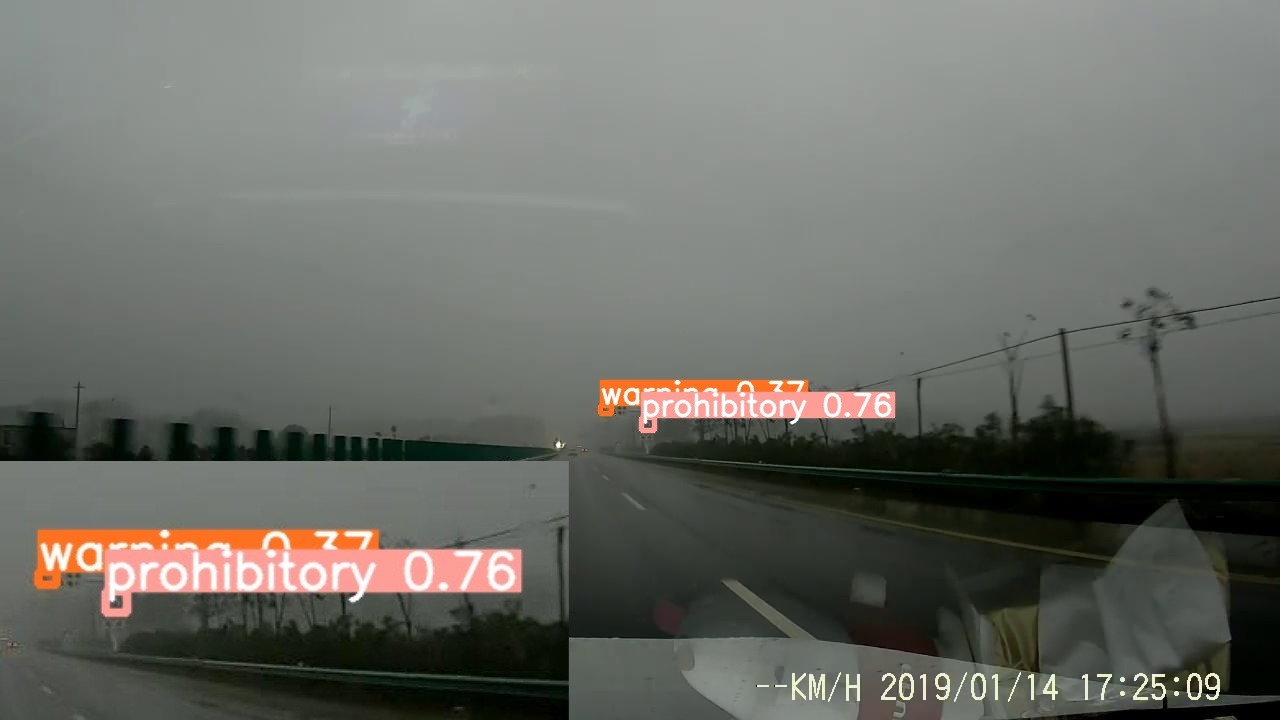}
              \vspace{0.4em}
              \includegraphics[width=2.6in]{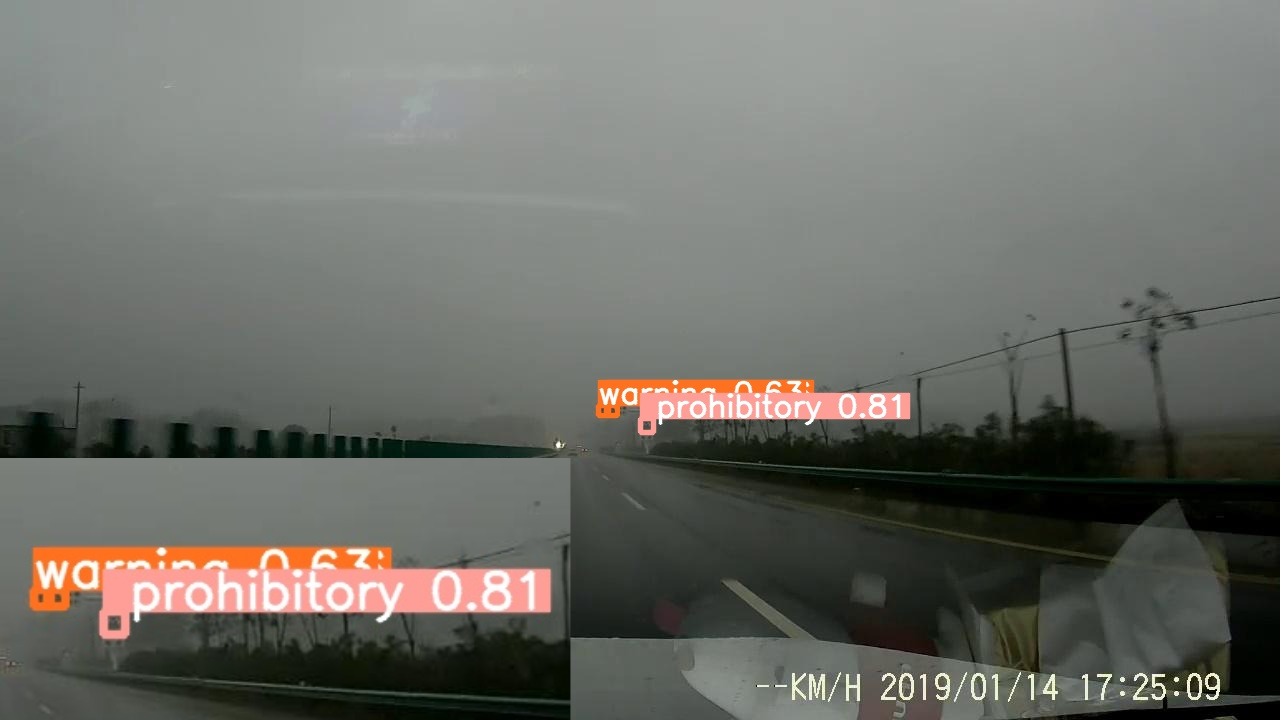}
              \label{fig_first_case}}
\hfil
\subfloat[{\fontsize{8pt}{8pt}\selectfont Snowy day test comparison results}]{\includegraphics[width=2.6in]{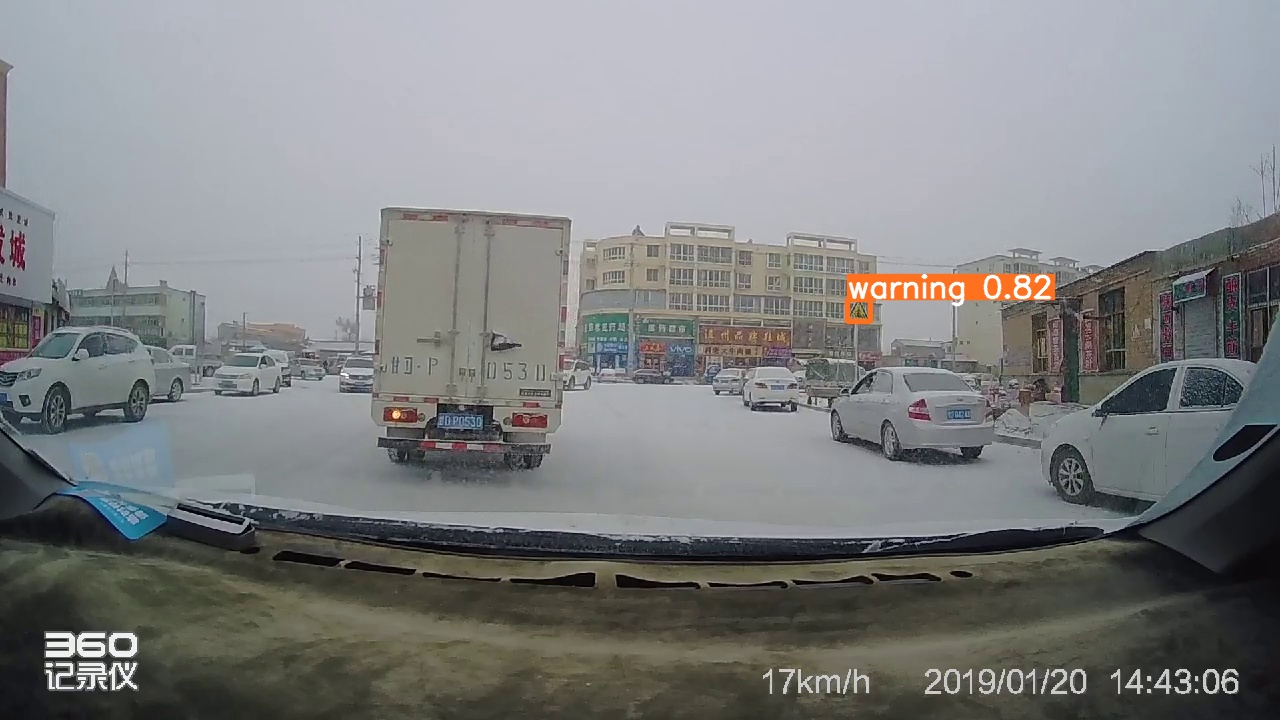}
              \vspace{0.4em}
              \includegraphics[width=2.6in]{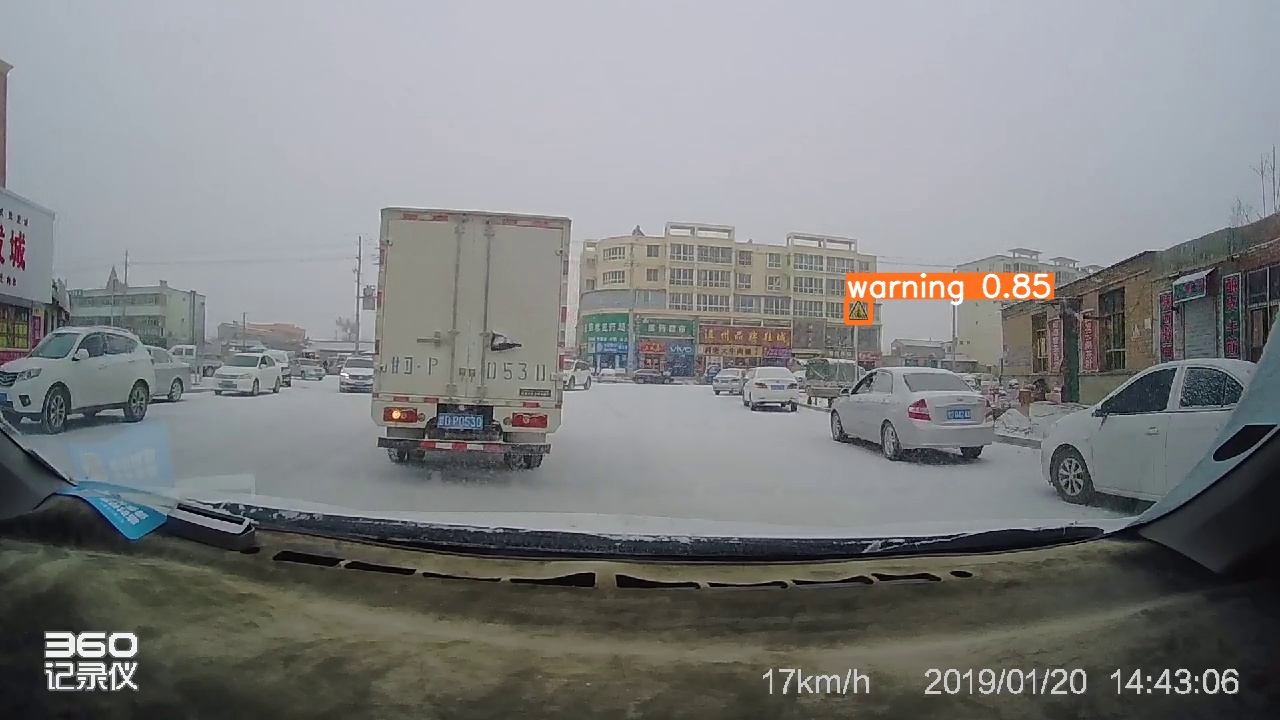}
              \label{fig_first_case}}
\hfil
\subfloat[{\fontsize{8pt}{8pt}\selectfont Night test comparison results}]{\includegraphics[width=2.6in]{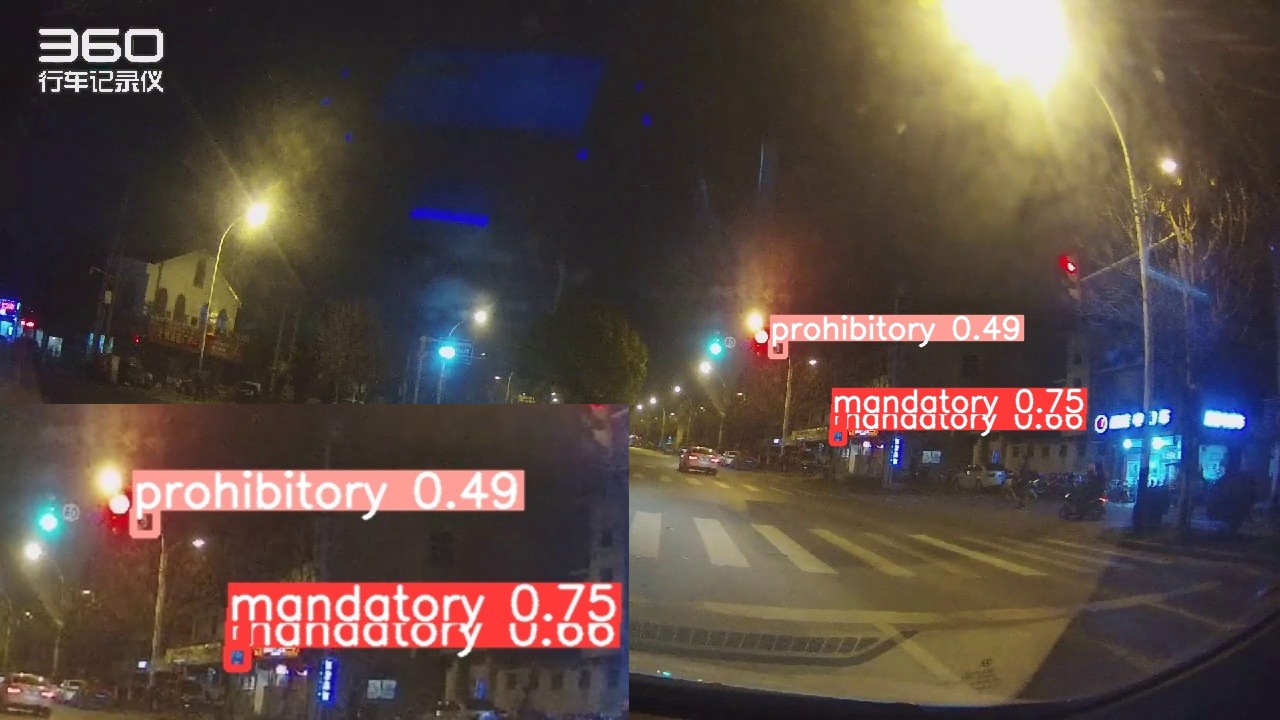}
              \vspace{0.4em}
              \includegraphics[width=2.6in]{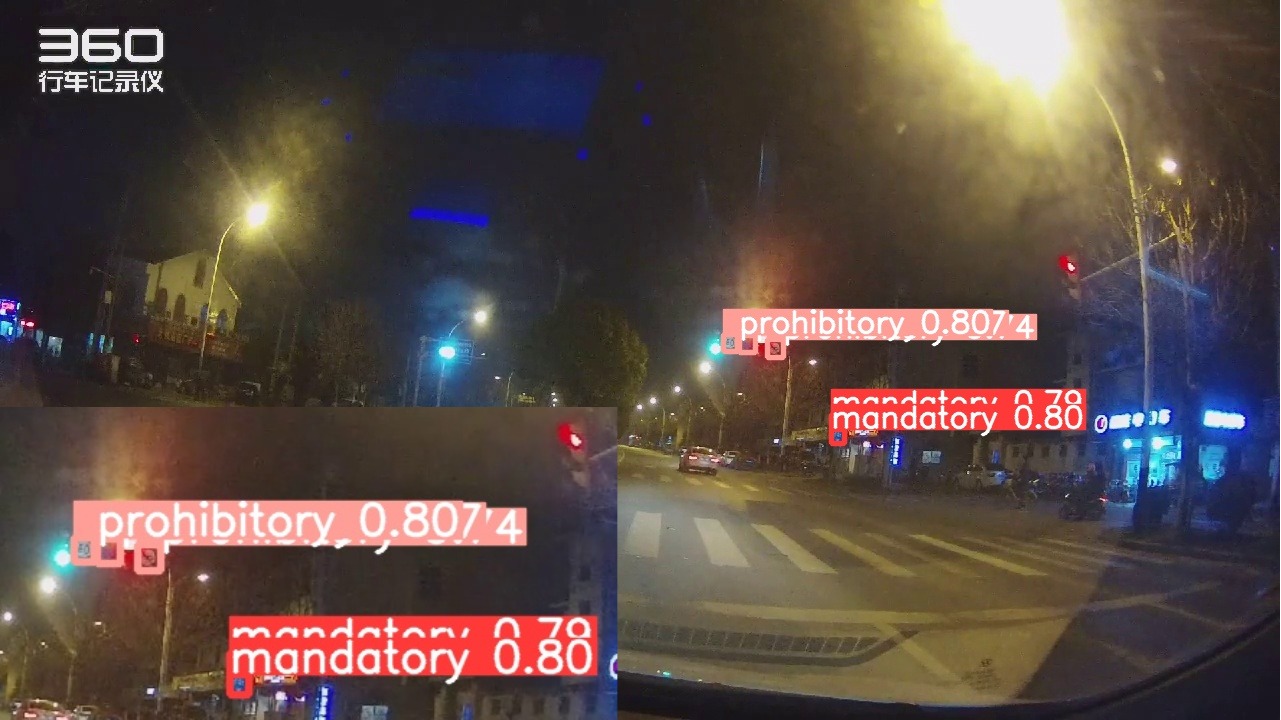}
              \label{fig_first_case}}
\caption{Detection examples under different environmental conditions on the CCTSDB2021 dataset. The first column displays the detection baseline. The second column displays the detection of the proposed YOLO-CCA, where all the traffic signs in the second column are accurately detected and the recognition confidence of the traffic signs is also better than the baseline model.}
\label{fig_sim}
\end{figure*}
\subsection{Ablation Analysis}
To verify the effectiveness of the designed LCFE, GCFC and Transformer-based CCA modules, ablation experiments are designed for verification. Experiments are conducted on the TT100K dataset using the same parameters, and the results of the experiments are shown in Table \uppercase\expandafter{\romannumeral2}. Experiment B is implemented by removing the LCFE and GCFC from the CCA, and retaining only the TransformerBlock. As can be seen in Table \uppercase\expandafter{\romannumeral2}, Experiment B improves the mAP@.5 by 1.6\% and reduces the amount of parameters by 4.5M by using the Transformer. After adding the LCFE module to Experiment B (Experiment C), mAP@.5 improves by 1.8\%, proving that the local context information provided by the LCFE improves the accuracy of object detection. Experiment D is implemented by keeping the GCFC in the CCA, and the position of the LCM is replaced by a convolution. Experiment D improves mAP@.5 by 0.5\% on Experiment B. So it is proved that the key position extracted by GCFC can help to detect the object that needs to be detected. Experiment E is our final improved model, with 3.9\% improvement in mAP@.5 and 3.4\% improvement in mAP@.5:.95 over the benchmark model, and 2.7M reduction in the amount of parameters, which proves the effectiveness of each module designed in this paper.

\subsection{Comparison of Methods}
\begin{table*}[t]
\caption {the quantitative comparison of different methods on CCTSDB2021 dataset.}
\label{tab:table1}
\centering
\begin{tabular}{l p{55pt}<{\centering}p{55 pt}<{\centering}p{55pt}<{\centering}p{55 pt}<{\centering}p{55 pt}<{\centering}p{55 pt}<{\centering}}
\hline
\textbf{Model} & \textbf{Params} & \textbf{FLOPS} & \textbf{P}     & \textbf{R}     & \textbf{mAP@.5} & \textbf{mAP@.5:.95} \\ \hline
\multicolumn{1}{l}{Faster-RCNN}    & 41.6M           & 211.5G         & 0.724          & 0.689          & 0.692           & 0.357               \\
\multicolumn{1}{l}{YOLOv3}         & 58.6M           & 155.3G         & 0.906          & 0.754          & 0.819           & 0.541               \\
\multicolumn{1}{l}{YOLOv3-SPP}     & 59.6M           & 156.1G         & 0.884          & 0.771          & 0.826           & 0.545               \\
\multicolumn{1}{l}{YOLOv5s}        & 6.7M            & 16.0G          & 0.901          & 0.736          & 0.807           & 0.525               \\
\multicolumn{1}{l}{YOLOv5l}        & 44.0M           & 108.3G         & 0.901          & 0.736          & 0.807           & 0.525               \\
\multicolumn{1}{l}{YOLOv8n}        & 2.8M   & \textcolor{blue}{8.2G}  & 0.914          & 0.674          & 0.770           & 0.490               \\
\multicolumn{1}{l}{YOLOv8s}       & 10.6M           & 28.7G          & 0.922 & 0.737          & 0.828           & 0.534               \\
\multicolumn{1}{l}{YOLOv8m}       & 24.6M           & 79.1G          & 0.882          & 0.770          & 0.841           & 0.561               \\
\multicolumn{1}{l}{YOLOv9}   & 29.8M & 118.2G & \textcolor{blue}{0.924} & 0.794 & 0.864 & 0.581 \\
\multicolumn{1}{l}{YOLOv10n} & \textcolor{blue}{2.6M}  & 8.4G   & 0.879 & 0.703 & 0.770  & 0.489 \\
\multicolumn{1}{l}{YOLOv10s} & 7.7M  & 24.9G  & 0.890 & 0.736 & 0.803 & 0.520 \\
\multicolumn{1}{l}{YOLOv10m} & 15.7M & 64.0G  & 0.897 & 0.757 & 0.822 & 0.547 \\
\multicolumn{1}{l}{YOLOv10b} & 19.4M & 98.8G  & 0.905 & 0.776 & 0.823 & 0.560 \\
\multicolumn{1}{l}{YOLOv10l} & 24.5M & 127.2G & 0.870 & 0.784 & 0.834 & 0.564  \\
\multicolumn{1}{l}{YOLOv7(baseline)}        & 35.4M           & 105.1G         & 0.918          & \textcolor{blue}{0.808} & 0.859           & 0.585               \\
\multicolumn{1}{l}{Ours}           & 33.6M           & 98.5G          & 0.920          & 0.802          & \textcolor{blue}{0.869}  & \textcolor{blue}{0.593}      \\ \hline
\end{tabular}
\end{table*}

\begin{figure*}[]
\centering
\begin{minipage}{\linewidth}
  \centering
\end{minipage}
\vspace{0.4em}
\begin{minipage}{\linewidth}
  \centering
  \includegraphics[width=2.3in]{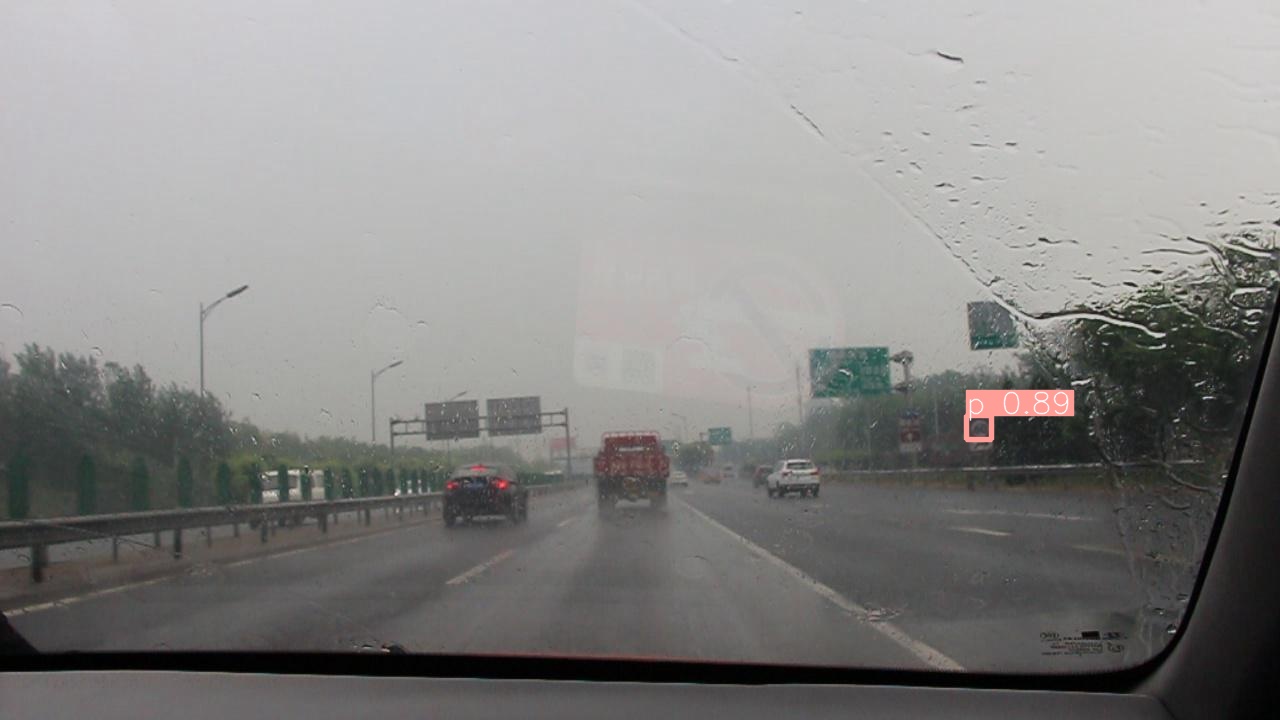}
  \includegraphics[width=2.3in]{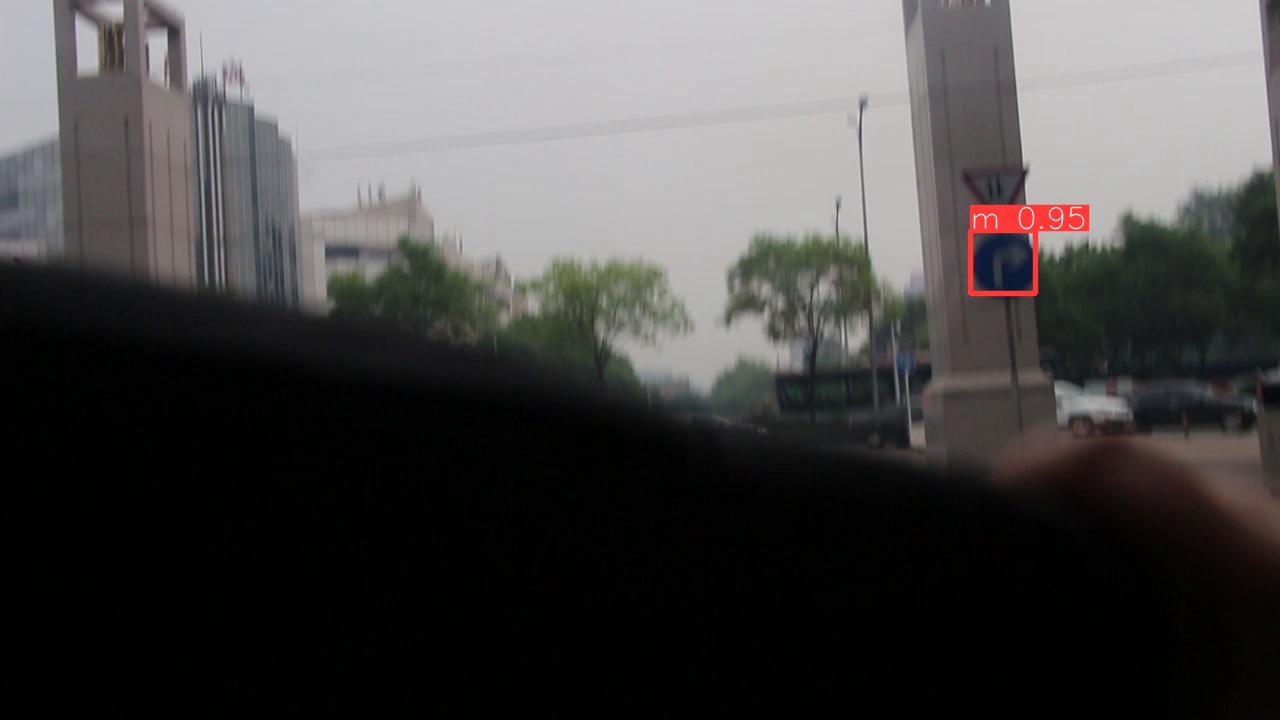}
  \includegraphics[width=2.3in]{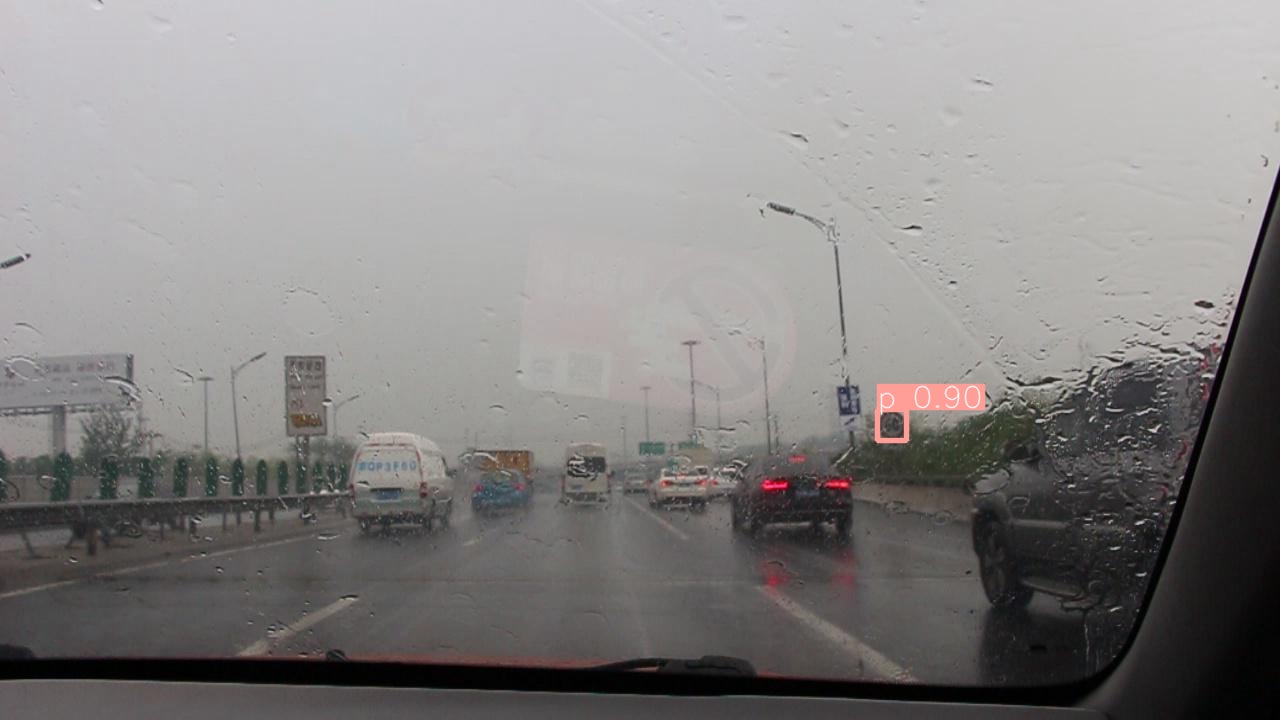}
\end{minipage}
\vspace{0.4em}
\begin{minipage}{\linewidth}
  \centering
  \includegraphics[width=2.3in]{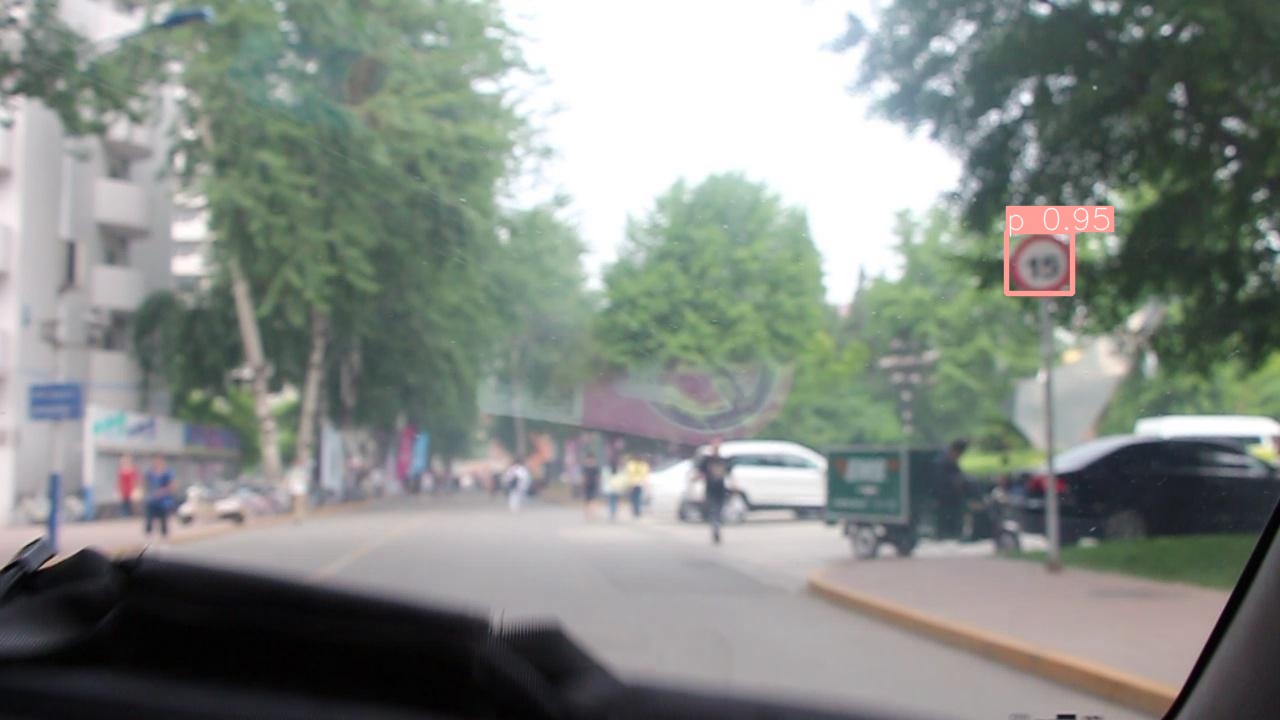}
  \includegraphics[width=2.3in]{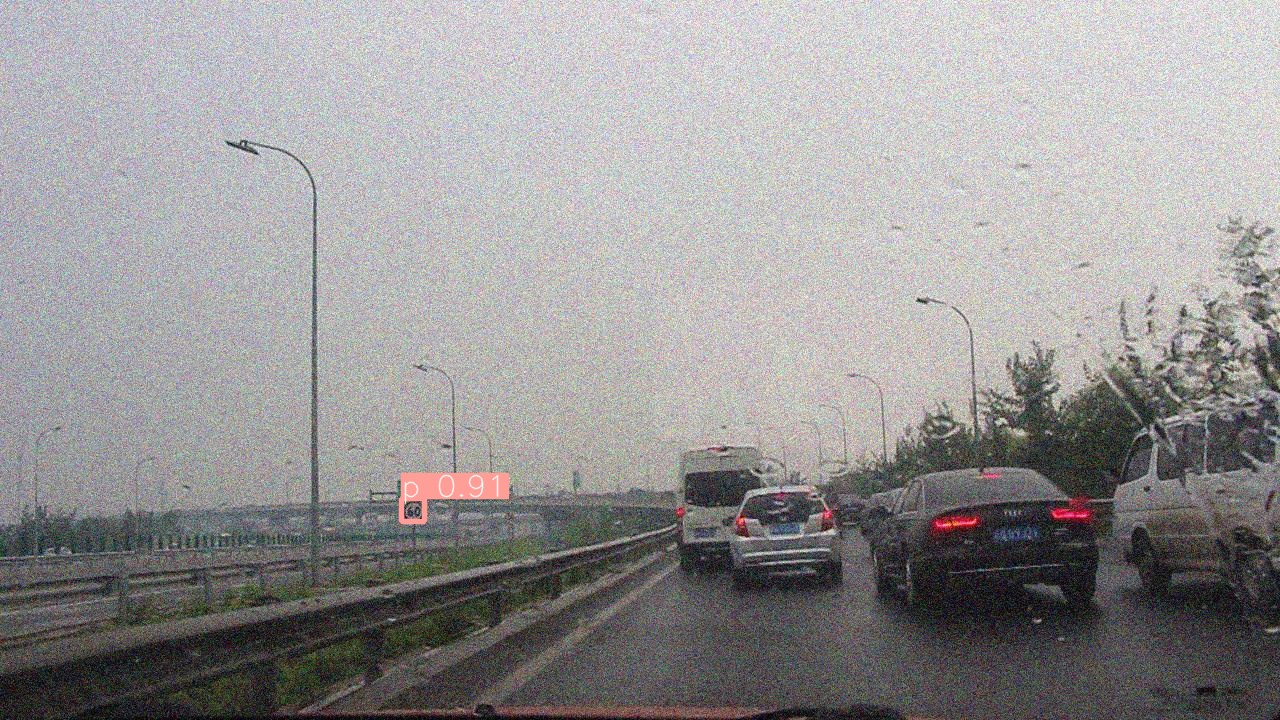}
  \includegraphics[width=2.3in]{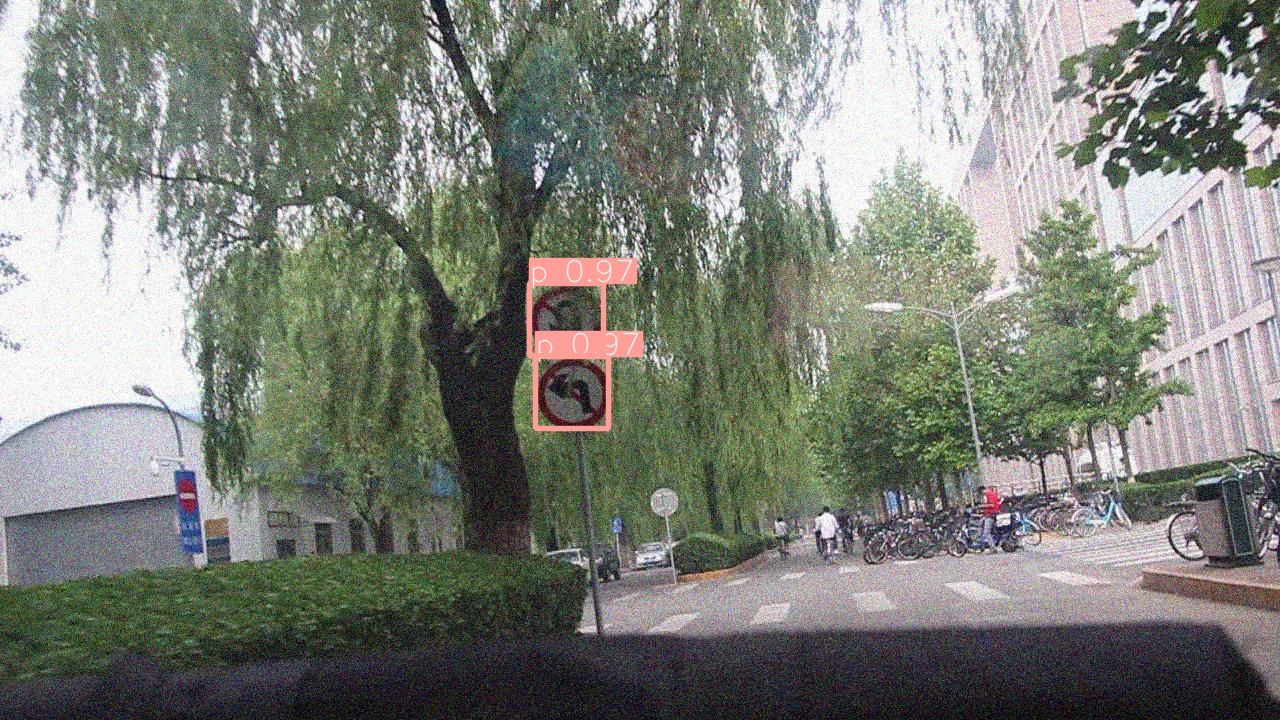}
\end{minipage}
\vspace{0.4em}
\begin{minipage}{\linewidth}
  \centering
  \includegraphics[width=2.3in]{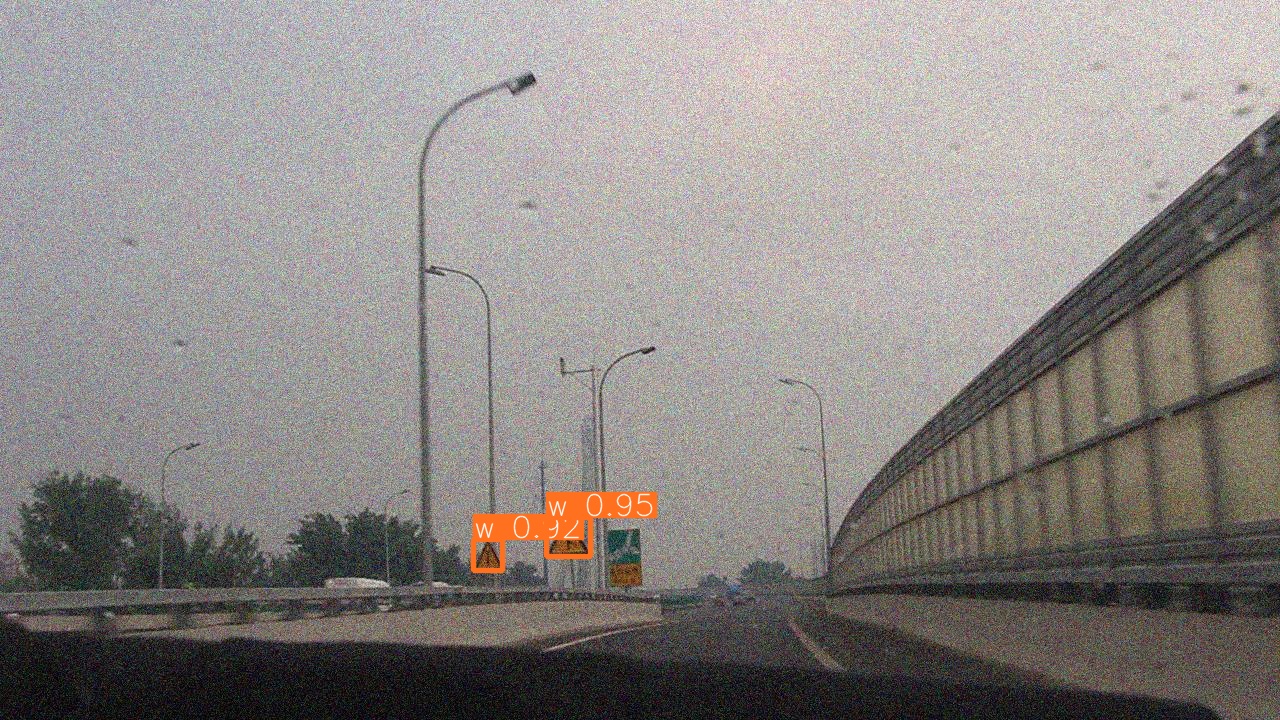}
  \includegraphics[width=2.3in]{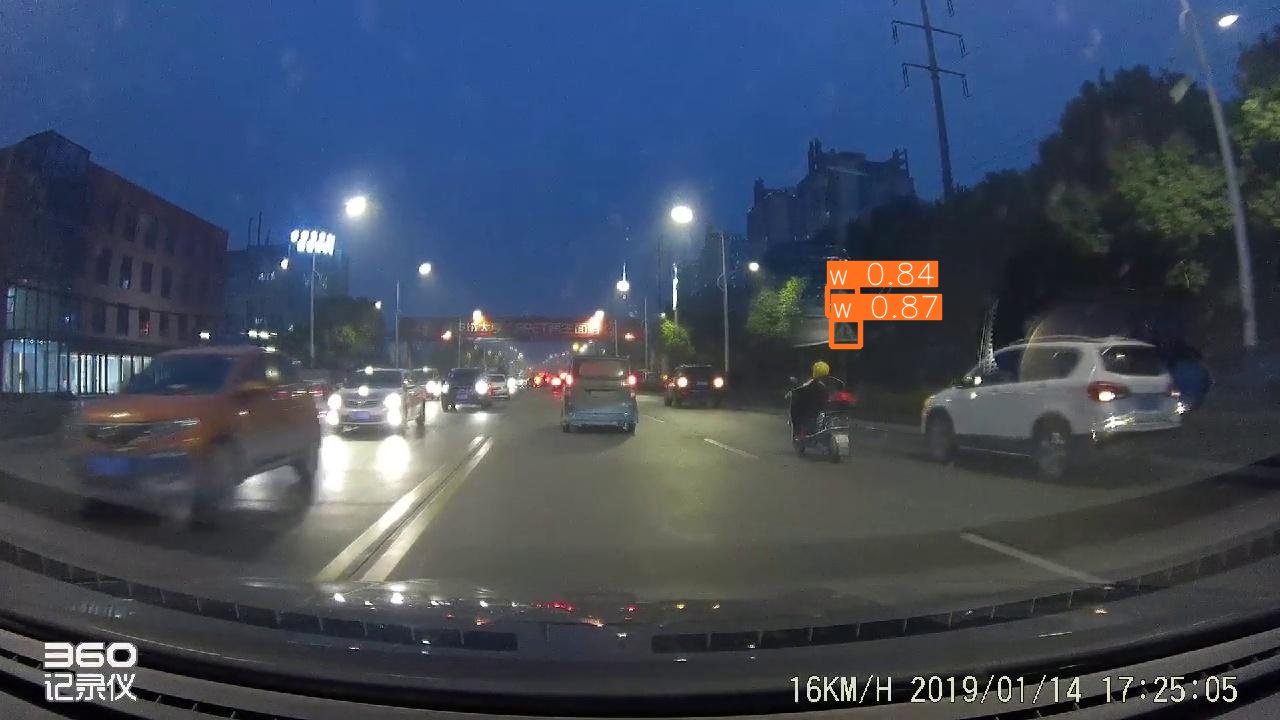}
  \includegraphics[width=2.3in]{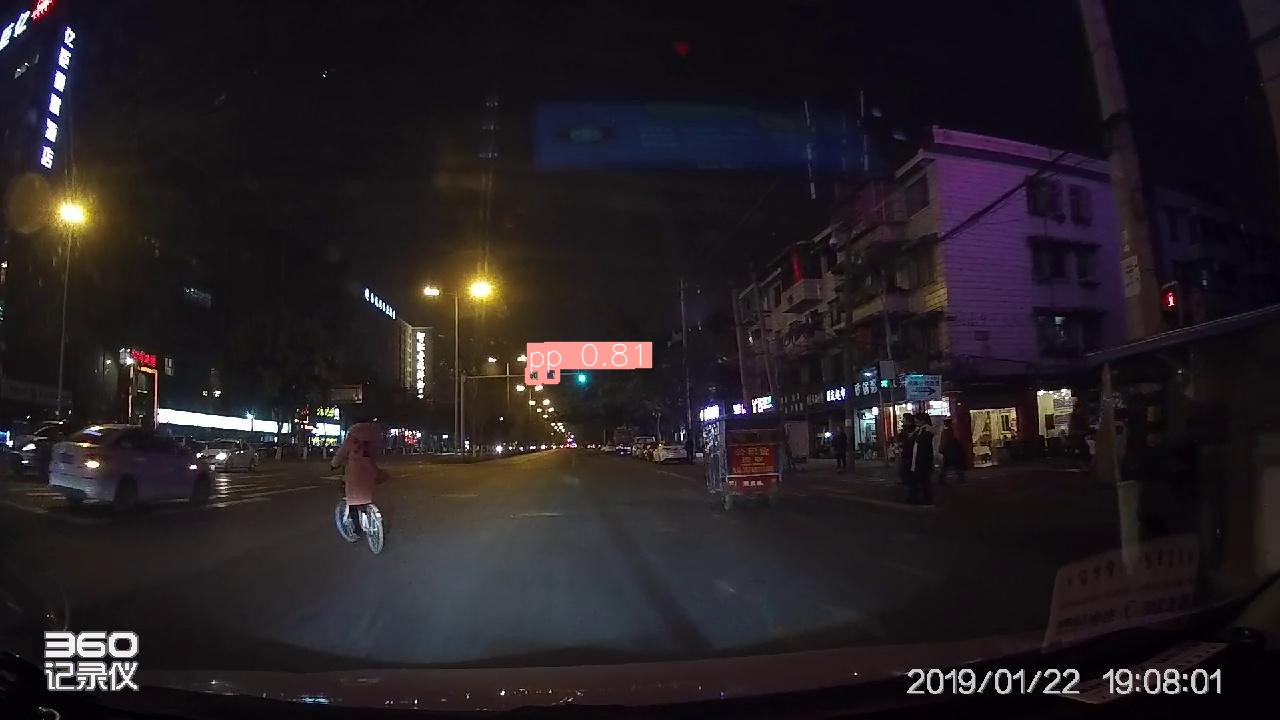}
\end{minipage}
\caption{Detection examples of complex traffic environments and challenging traffic sign scenes in the CCTSDB2021 dataset. The CCTSDB2021 dataset consists of three traffic sign categories: `m' representing mandatory signs, `p' representing prohibition signs, and `w' representing warning signs. }
\label{fig_8}
\end{figure*}

A comprehensive comparison of our method with other mainstream algorithms on the TT100K dataset and the experimental results are shown in Table \uppercase\expandafter{\romannumeral3}. Compared with the baseline model YOLOv7, mAP@.5 improves by 3.9\% and mAP@.5:.95 improves by 3.4\%. From the experimental results, the improvement of our method is more obvious, which proves that our proposed YOLO-CCA has better detection effect.

Fig. 8 shows the visualized detection results of this paper's method and YOLOv7 on the TT100K dataset. The first row of images is YOLOv7 and the second row is the method of this paper. For ease of viewing, the area containing the traffic sign is enlarged and placed below the image. In the first column of images, YOLOv7 misses the small-scale traffic signs located on distant poles, while our method accurately detects the smaller ‘pn’ traffic signs. As can be seen in the second column of images, both YOLOv7 and our method accurately detect traffic signs, but our method detects traffic signs with higher confidence scores. In the third column of images, YOLOv7 misdetects ‘pl40’, which is located below the traffic signal, as ‘p11’.  This is thanks to the local context and global context introduced in this paper, which improves the detection accuracy of traffic signs by utilizing the relevant information around the traffic signs and the features at some key locations in the image. The visualization results show that the method in this paper effectively improves the omission and misdetection of small-scale traffic signs, and can detect small-scale traffic signs more accurately.

To verify the performance of YOLO-CCA model and its robustness to other traffic sign datasets, YOLO-CCA is compared with other object detection methods on the CCTSDB2021 dataset. The comparison results are shown in Table \uppercase\expandafter{\romannumeral4}. YOLO-CCA achieves 86.9\% mAP@.5  and 59.3\% mAP@.5:.95 on the CCTSDB2021 dataset. Compared to the baseline model YOLOv7 mAP@.5 increases by one percentage point and mAP@.5:.95 by 0.8\%. From the Table \uppercase\expandafter{\romannumeral4}, it can also be seen that the mAP of our method is better than that of other methods, and the accuracy exceeds that of the state-of-the-art models and reaches the highest accuracy rate. The experiments show that the algorithm proposed in this paper is also robust on other traffic sign data.

Fig. 9 shows the comparison of traffic sign detection effect between YOLOv7 and YOLO-CCA in different environments, with the detection effect of YOLOv7 on the left and that of YOLO-CCA on the right. From the three figures in Fig. 9(a)-Fig. 9(c), it can be seen that YOLO-CCA can detect the traffic signs that YOLOv7 fails to recognize in abnormal weather such as foggy, rainy, and snowy days, and the confidence  of YOLO-CCA in recognizing the traffic signs is better than that of YOLOv7. Fig. 9(d) shows the traffic sign detection in dark environment, from which it can be seen that YOLO-CCA can accurately detect traffic signs in dark condition. YOLOv7 only detects three objects at the intersection. This indicates that YOLO-CCA performs well in detecting objects under different lighting and weather conditions.

To further validate the effectiveness of the YOLO-CCA model in handling complex scenarios, this paper introduces more challenging detection samples including adverse weather conditions, nighttime environments, and challenging situations such as occlusions and blur. The detection results shown in Fig. 10 demonstrate the capability of our proposed YOLO-CCA model to maintain excellent detection accuracy even when faced with these complex conditions, showcasing its robust adaptability. 

\begin{figure}[t]
\centering
\begin{minipage}{\linewidth}
  \centering
\end{minipage}
\vspace{0.4em}
\begin{minipage}{\linewidth}
  \centering
  \includegraphics[width=3.in]{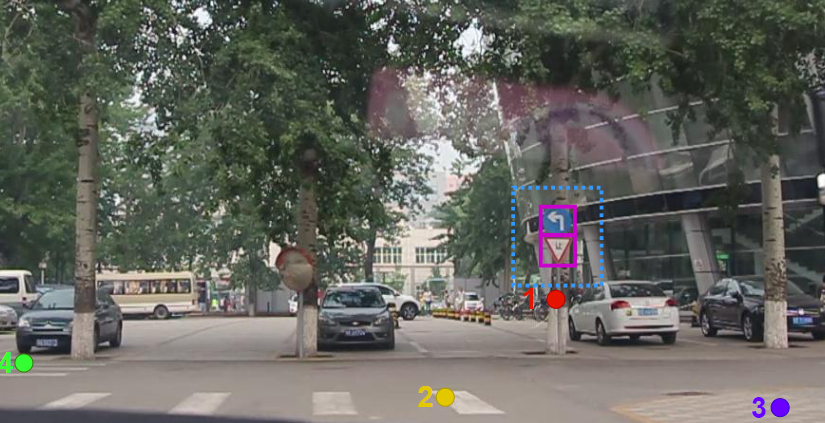}

\end{minipage}
\vspace{0.4em}
\begin{minipage}{\linewidth}
  \centering
  \includegraphics[width=3.in]{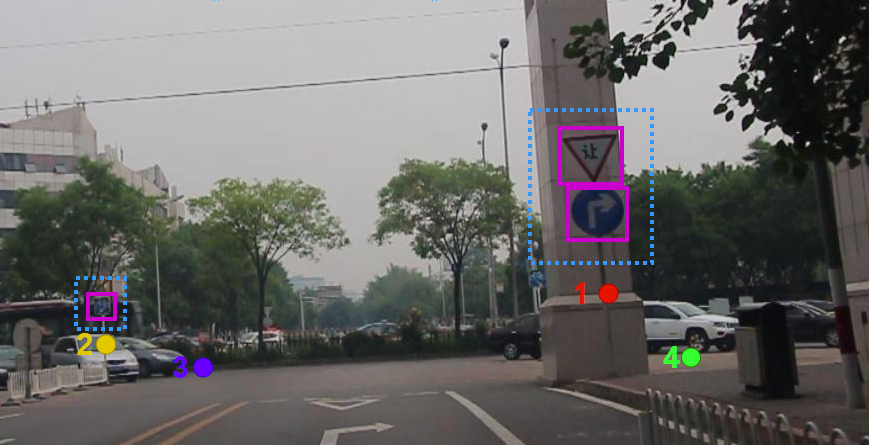}
\end{minipage}
\caption{The resulting visualization of the local context and global context collected by CCA. The purple boxes represent the bounding boxes, the blue dashed rectangles indicate the locally concentrated context, and the colored dots depict the positions of the summarized key global context. }
\label{fig_8}
\end{figure}

\subsection{Visualize Local and Global Contexts}
In addition to ablation analysis and comparative analysis, we also attempted to visualize the context information collected by the proposed CCA.

Fig. 11 presents the visualization results. In this figure, we primarily showcase the positions of the concentrated local context (blue dashed rectangles) and the aggregated key global context (colored dots). For traffic sign targets that are challenging to detect due to factors like occlusion, blur, deformation, the locally concentrated context around the bounding box is crucial. From the image, we can also observe that the summarized key global context is typically located within meaningful instances. For instance, in the top image, the learned key global context is distributed around the traffic sign pole, zebra crossing, and turn, which is reasonable for detecting traffic sign images. Meanwhile, the surrounding vehicles and large trees are less significant, which aligns with common sense. These results clearly indicate that CCA can effectively model the relationship between global context and locally concentrated context, thus aiding in achieving efficient detection based on context and modeling relationships.

\section{Conclusion and Future Work}
In this paper, we have proposed a context-based traffic sign detection algorithm, YOLO-CCA, to address the challenges associated with small traffic sign, limited feature information, and low detection accuracy. We have extracted both local and global contextual information and utilized the Transformer to fuse these two types of contexts, resulting in improved multi-level feature fusion results. Our proposed CCA module utilizes relevant information surrounding the traffic signs and key positional features of the image scene to enhance the feature fusion capability of the network. It exhibits improved performance in handling complex scenarios. We have conducted comparative experiments on the TT100K dataset, and the results demonstrate that our model achieves a significant improvement in detection accuracy, with an mAP of 92.1\%, which is 3.9\% higher than the baseline network. The detection performance for small objects is significantly improved, with a notable reduction in both false positives and false negatives. We also have tested YOLO-CCA on the CCTSDB2021 dataset, where it achieves an mAP of 86.9\%, which is 1\% higher than the baseline network, further validating the effectiveness of our approach. 

 We utilize two types of contextual information solely within the feature fusion network to optimize the fusion results of multi-level features.  However, these contextual cues have not yet been integrated into the backbone network.  In the future, our focus will be on exploring effective methods to incorporate these contextual information  into the backbone architecture of the object detection network, thereby enhancing the feature extraction and representation capabilities of the backbone network.

\bibliographystyle{IEEEtran}
\bibliography{re}

\vfill
\end{document}